\documentclass[a4paper,11pt]{article}

\usepackage{jheppub} 

\usepackage{graphicx}
\usepackage{amsmath}
\usepackage{enumitem}

\usepackage{epsfig}

\usepackage{subfigure}
\usepackage{longtable}
\usepackage{times}

\title{Bondi-Hoyle-Lyttleton Accretion around the Rotating Hairy Horndeski Black Hole}

\author{O. Donmez}
\affiliation{College of Engineering and Technology, American
  University of the Middle East, Egaila 54200, Kuwait}

\emailAdd{orhan.donmez@aum.edu.kw}

\abstract{Modeling of the shock cone formed around a stationary, hairy Horndeski black hole
  with Bondi-Hoyle-Lyttleton (BHL) accretion has been conducted.
  We model the dynamical changes of the shock cone resulting from the interaction of matter
  with the Horndeski black hole, where the scalar field and spacetime have a strong interaction.
  The effects of the scalar hair, the black hole rotation parameter, and the impacts of the asymptotic
  speed have been examined, revealing the influence of these parameters on the shock cone and the
  trapped QPO modes within the cone. Numerical calculations have shown that the hair parameter
  significantly affects the formation of the shock cone. As the absolute value of the hair parameter
  increases, the matter in the region of the shock cone is observed to move away from the black hole horizon.
  The rate of matter expulsion increases as $h/M$ changes.
  After $h/M<-0.6$, a visible change in the physical structure of the shock cone occurs,
  ultimately leading to the complete removal out of the shock cone.
  On the other hand, it has
  been revealed that the asymptotic speed significantly affects the formation of the
  shock cone. As $h/M$ increases in the negative direction and the asymptotic speed increases,
  the stagnation point moves closer to the black hole  horizon. When the value of the hair
  parameter changes, the rest-mass density of the matter inside the cone decreases, whereas the
  opposite is observed with the asymptotic speed. Additionally, the formed shock cone has excited QPO modes.
  The deformation of the cone due to the hair parameter has led to a change or complete
  disappearance of the QPOs.
  Meanwhile, at asymptotic speeds of $V_{\infty}/c< 0.4$, all fundamental
  frequency modes are formed, while at $V_{\infty}/c=0.4$, only the azimuthal mode is excited,
  and $1:2:3:4:...$ resonance conditions occur. No QPOs have formed at $V_{\infty}/c = 0.6$.
  The results obtained from numerical calculations have been compared with theoretical studies
  for $M87^*$, and it has been observed that the possible values of $h/M$ found in the
  numerical simulations are consistent
  with the theory. Additionally, the results have been compared with those for the GRS 1915+105
  black hole, and the hair parameters corresponding to the observed frequencies have been
  determined.
}

\keywords{
numerical relativity, rotating black hole, alternative gravities, Bondi-Hoyle-Lyttleton, QPOs}

\begin{document} 
\maketitle
\flushbottom


\section{Introduction}
\label{Introduction}

Due to various physical reasons, mass accretion  in Active Galactic Nuclei (AGNs) and X-ray binary
systems causes intense emission of electromagnetic radiation across a wide range of frequencies
due to their interactions with  the stellar and  massive black holes they harbor at their centers.
The observation of X-rays emitted as a result of interactions between accreted matter and massive,
as well as stellar-mass black holes has yielded significant scientific insights.
One of the most notable observational achievements in recent times is the unveiling of the shadows of
black holes at the centers of the  the $M87$ \citep{Akiyama1, Akiyama2} and Milky Way
galaxies \citep{Akiyama3, Akiyama4}.
Conversely, the observation of Quasi-Periodic Oscillation (QPO) frequencies  in X-ray binaries
\citep{Remillard1999ApJ, Naik2000A&A, Zhang2020MNRAS, Ingram2019, Reynolds1, Smith2021ApJ, Jin2023ApJ}
over decades is crucial for
unveiling the mysteries of the universe and understanding the physical properties of black holes.
These observations have illuminated various astronomical phenomena while
facilitating the calculation of physical properties  such as mass and rotation parameters of the black holes
across different ranges.
Therefore, understanding the accretion disks formed by the falling  matter toward the black holes and
the resulting physical mechanisms is important in gravitational wave astronomy.

There are various models related to mass accretion around compact objects, with one of the most
significant being BHL accretion. Initially defined by Bondi and Hoyle and
further developed by Lyttleton \citep{Bondi1, Bondi1952MNRAS}, this phenomenon occurs as an object
moves through a medium, forming a disk where matter is gravitationally pulled toward the other side of
the compact object. Numerous numerical results on BHL accretion onto black holes exist in the literature,
revealing its potential to form accretion disks, shock cones, increase black hole mass through continuous
matter accretion, form binary star systems, and produce stars in the interstellar medium.
The historical development of BHL accretion and its formulations in both Newtonian and
relativistic frameworks can be found in Refs.\cite{Edgar1,Luciano2015GReGr}.
Relativistic simulations play a crucial role in revealing the physical properties of the
black holes and in explaining $X-$ray data observed by detectors, particularly in regions
very close to the black hole horizon  where gravitational forces are dominant.
These simulations yield accurate numerical solutions in areas where gravitational effects
are significant, making them essential for understanding  phenomena occurring near the black holes.
In regions of very strong gravitational forces,
numerous simulations related to the Schwarzschild and the Kerr black holes
have been conducted to understand the BHL mechanism and contribute to observational results.
These simulations are critical for interpreting the complex dynamics near the black holes,
where the gravitational effects significantly influence the behavior of  the accreting matter
\citep{Donmez6, Penner1, Zanotti1, Donmez5, Penner2, LoraClavijo1, Koyuncu1,
  LoraClavijo2, CruzOsorio2020ApJ,CruzOsorio2023JCAP}.
In recent years, using alternative theories of gravity, the dynamic structure of the
shock cone in regions with very strong gravitational forces has been examined
\citep{Donmez3, Donmez_EGB_Rot, Donmez2024Univ}. It has
been revealed how the shock cone changes depending on various parameters. Simultaneously,
the QPO frequencies generated in such systems have been numerically calculated,
and their compatibility with observations has been investigated.

Modified gravity has recently gained popularity in numerical relativity
due to the limitations of well-known black hole models, such as the Kerr solution, in explaining certain
observational results and phenomena like dark matter and dark energy.
Alternative theories offer potential solutions to these issues.
Alternative theories offer potential solutions to these issues. One such alternative theory is
Horndeski, which allows for the coupling of the scalar field with spacetime by defining a 4-dimensional
spacetime matrix and is free of Ostrogradski instability \citep{Horndeski1974IJTP}. Consequently,
the Horndeski black hole is characterized by the scalar hair parameter that describes the scalar field.
Both non-rotating and rotating Horndeski black holes  have been theoretically
studied in the literature, yielding significant progress in solving known astrophysical problems
and explaining observational results.
These efforts include providing solutions to phenomena such as strong gravitational lensing \citep{Kumar2022EPJC},
cosmological studies \citep{Kubota2023PhRvD}, observed black hole shadows in M87 and the Milky
Way galaxy \citep{Afrin2022ApJ}, QPO behaviors \citep{Rayimbaev2023EPJC},
detection of the photon ring \citep{Gao2023EPJC, Ghosh2023ApJ, Wang2023PhRvD}, the impact of tilted accretion disks on
observations \citep{Hu2023arXiv}, and luminosity in thin disks \citep{Heydari-Fard2023arXiv}.

Studying the accretion disks around the massive black holes by incorporating the effects of scalar
fields may help reveal interactions between baryonic matter and dark matter. Because the scalar field is considered one of the candidates that might define dark matter, revealing the impact of the scalar field on the disk formed around the black hole may allow some predictions to be made about the nature of dark matter \citep{Magana2012JPhCS, TellezTovar2022PhRvD, CruzOsorio2023JCAP, Gomez2024PhRvD}. Similar to Horndeski gravity, the interaction of scalar fields with gravity, through the hair parameter,
could play a significant role at both galactic \citep{Matos2000PhRvD}
and cosmic scales \citep{Matos2000CQGra} in understanding the physical properties of the black holes,
and unveiling the mystery of dark matter.
Horndeski gravity encompasses  a rich content and exhibits parallelism with the observational results
in cases of  the accelerating expansion of the universe and modifications of gravitational interactions.
The scalar hair parameter  in Horndeski gravity
affects the disk structure near the black hole horizon, the mass accretion process,
the emission of gravitational waves, and the shadow of the black hole observed by the Event
Horizon Telescope \citep{Afrin2022ApJ, Ghosh2023ApJ}.

In this paper, for the first time, we reveal the dynamic structure of the accretion disk
and shock cone around the  rotating Horndeski black hole in case of the BHL accretion
by numerically solving the General Relativistic Hydrodynamic (GRH) equations. To achieve this,
we direct  matter towards the rotating black hole from the upstream region of the computational
domain and examine the effect of the Horndeski gravity scalar field parameter $h/M$,
namely the scalar hair parameter, on the shock cone, particularly in the strong gravitational fields
very close the black hole horizon. We uncover the impact of the hair, asymptotic velocity of the matter
injected from outer boundary, and the black hole rotation
parameters on the structure of the shock cone, the mass accretion rate, and the produced QPO frequencies
inside the shock cone. We compare the numerical solutions computed from the Horndeski black hole 
with the Kerr solution and attempt to explain some of the observational data.

The organization of the paper is as follows. Section \ref{Equations} provides the GRH equations
and the rotating Horndeski black hole metric. In Section \ref{InitBoun}, the initial conditions,
the range of the hair parameter that is used in numerical simulations, and the boundary conditions
are explained.
The formation mechanisms of the shock cone at different rotation
parameters and scalar field parameter values, how the cone disappears in some extreme cases,
and the significant changes in the dynamic structure of the shock cone due to the
asymptotic speed with the strong scalar field parameter are described in Section \ref{Result1}.
Section \ref{Possible_QPO} numerically reveals the possible QPO frequencies for all models
through the power spectrum density analysis, providing detailed information on how QPOs are excited
or disappear under different dynamic conditions of the shock cone. The compatibility of the
obtained models  of the shock cone structures and QPO situations with theoretical studies related
to the $M87^*$ black hole using the possible hair parameters is presented in Section \ref{M87}.
In Section \ref{Horndeski_Mass_QPO}, observational results obtained from the GRS 1915 + 105
source are compared with our numerical results, and some QPO and mass limitations are made.
Finally, the findings are summarized in Section \ref{Conclusion}. Throughout the paper, the geometrized
units are used, meaning $G=c=1$. Therefore, length and time quantities are expressed in terms
of the mass of the black hole.


\section{Equations}
\label{Equations}

\subsection{General Relativistic Hydrodynamic Equations}
\label{GRHE1}

To reveal the physical characteristics of the disk around the black holes and understand the observed
data in the strong gravitational region, the General Relativistic Hydrodynamic (GRH)
equations are written in conservation form using the 3+1 formalism\citep{Font2000LRR, Donmez1}.
Ignoring the effects of magnetic fields and viscosity, the GRH  equations on the equatorial
plane  for a perfect fluid stress-energy tensor are,

\begin{eqnarray}
  \frac{\partial U}{\partial t} + \frac{\partial F^r}{\partial r} + \frac{\partial F^{\phi}}{\partial \phi}
  = S,
\label{GRHE2}
\end{eqnarray}

\noindent
where  $U$, $F^r$, $F^{\phi}$, and $S$ are conserved variables, fluxes in  the $r$ and $\phi$ directions,
and the source,   respectively. The conserved variables,  fluxes and sources are written in terms of primitive
variables, 3-metric, and other variables defined later as follows.

\begin{eqnarray}
  U =
  \begin{pmatrix}
    D \\
    S_j \\
    \tau
  \end{pmatrix}
  =
  \begin{pmatrix}
    \sqrt{\gamma}W\rho \\
    \sqrt{\gamma}\xi \rho W^2 v_j\\
    \sqrt{\gamma}(\xi \rho W^2 - P - W \rho)
    \end{pmatrix},
\label{GREq5}
\end{eqnarray}

\noindent and fluxes are

\begin{eqnarray}
  \vec{F}^i =
  \begin{pmatrix}
    \alpha\left(v^i - \frac{1}{\alpha\beta^i}\right)D \\
    \alpha\left(\left(v^i - \frac{1}{\alpha\beta^i}\right)S_j + \sqrt{\gamma}P\delta^i_j\right)\\
    \alpha\left(\left(v^i - \frac{1}{\alpha\beta^i}\right)\tau  + \sqrt{\gamma}P v^i\right)
    \end{pmatrix},
\label{GREq6}
\end{eqnarray}

\noindent and

\begin{eqnarray}
  \vec{S} =
  \begin{pmatrix}
    0 \\
    \alpha\sqrt{\gamma}T^{ab}g_{bc}\Gamma^c_{aj} \\
    \alpha\sqrt{\gamma}\left(T^{a0}\partial_{a}\alpha - \alpha T^{ab}\Gamma^0_{ab}\right)
   \end{pmatrix},
\label{GREq7}
\end{eqnarray}

\noindent
where $\xi = 1 + \epsilon + P/\rho$ is  the enthalpy, $\Gamma^c_{ab}$ represents the Christoffel symbol, 
$W = (1 - \gamma_{i,j}v^i v^j)^{1/2}$ is the Lorentz factor, $v^i = u^i/W + \beta^i$
donates the three-velocity of the fluid. $\xi$ , $\rho$, $\epsilon$, $\gamma_{i,j}$, $g^{ab}$, $\gamma$,
$u^{a}$, $p$, $\alpha$, and $\beta^i$
are the specific enthalpy, the rest-mass density, the internal energy, 3-metric,
the four-metric of the curved space-time, the determinant of three-metric, $4-$ velocity of the fluid,
and the fluid pressure,  the lapse function, and shift vector,
respectively. While the indices $i$, $j$, and $k$ go from $1$ to $3$,
$a$, $b$, and $c$ range from $0$ to $3$.

In this paper, Equation \ref{GRHE2} is solved numerically to reveal the dynamic structure of the shock cone generated by matter falling towards the black hole in a strong gravitational field. To achieve this, initial values consistent with BHL accretion are defined for the primitive variables. Using these initial values, fluxes are first calculated in the equatorial plane. These fluxes are then used in Riemann solvers to numerically determine the relativistic conservative variables at the next time step. Subsequently, primitive variables are calculated from the computed conservative variables \citep{Donmez1, Donmez2, Donmez6}. At this point, the flow speed of the matter must not exceed the speed of light. This situation is known as the "superluminalities" problem. To avoid superluminalities, the following measures are applied in numerical simulations. The Courant condition is used to prevent matter from moving faster than the speed of light within each grid cell. If the matter reaches a speed that could exceed the speed of light, the Courant condition reduces the time step to prevent this situation \citep{Donmez1, Donmez2}. Thus, for each model, the time intervals used over approximately 4 million time steps in numerical simulations are controlled by the Courant condition. Additionally, high-resolution shock-capturing methods and Riemann solvers are used in numerical solutions \citep{Landau1959, Luciano2015GReGr}. These methods specifically prevent superluminal conditions that could occur in shock waves within strong gravitational fields. Moreover, the boundary conditions used also ensure that the speed of the matter remains below the speed of light.


\subsection{Rotating Black Hole Space-Time Metric in Horndeski gravity}
\label{Horndeski_gravity}

By solving the GRH equations numerically, we reveal the
physical structure of the disk around a hairy black hole and its QPO behaviors.
For this purpose, in this section, we define the stationary spacetime metric, lapse function,
and shift vector of the Horndeski black hole.

  As a result of a more detailed examination of Horndeski gravity \citep{Horndeski1974IJTP}, it was shown that shift-symmetric Horndeski and beyond Horndeski theories can give rise to static and asymptotically flat black holes \citep{Babichev2017JCAP}. However, it was emphasized that the scalar field in this case must have a hair parameter. Theoretical studies concluded that the Lagrangian of Horndeski theory, taking this scalar field into account, is as follows \citep{Babichev2017JCAP}.

\begin{eqnarray}
    S =\int \sqrt{-g} \left( Q_2(\chi) + Q_3(\chi)\square\phi + Q_4(\chi)R + Q_{4,\chi}\left[( \square \phi)^2 -
      (\nabla^{\mu}\nabla^{\nu}\phi)(\nabla_{\mu}\nabla_{\nu}\phi)\right]+ \right.  \nonumber \\
   Q_5(\chi)G_{\mu\nu}\nabla^{\mu}\nabla^{\nu}\phi - \frac{1}{6}Q_{5,\chi}\left[( \square \phi)^3-3(\square\phi)(\nabla^{\mu}\nabla^{\nu}\phi)(\nabla_{\mu}\nabla_{\nu}\phi) + \right. \nonumber \\
   \left.   \left.  2(\nabla_{\mu}\nabla_{\nu}\phi)(\nabla^{\nu}\nabla^{\gamma}\phi)(\nabla_{\gamma}\nabla^{\mu})\right]\right)d^4x,  
    \label{Hornext1}
  \end{eqnarray}

\noindent
where $\chi=-\partial^{\mu}\phi \partial_{\mu}\phi/2$ is canonical kinetic term and $Q_2$, $Q_3$, $Q_4$, $Q_5$ are arbitrary functions for scalar field $\phi$.    $Q_{4,\chi}$ stands for $\partial Q_{4}(\chi)/\partial \chi$, $G_{\mu \nu}$ is the Einstein tensor, $R$ is the Ricci scalar, $(\nabla_{\mu}\nabla_{\nu}\phi)^2 =\nabla_{\mu}\nabla_{\nu}\phi \nabla^{\nu}\nabla^{\mu}\phi$, and $(\nabla_{\mu}\nabla_{\nu}\phi)^3 =\nabla_{\mu}\nabla_{\nu}\phi \nabla^{\nu}\nabla^{\rho}\phi \nabla_{\rho}\nabla^{\mu}\phi$. Using the restriction of scalar field  as $\phi = \phi(r)$ in order satisfy the static and spherically symmetric spacetime and other restriction given in Ref.\citep{Babichev2017JCAP, Esteban2021arXiv, Kumar2022EPJC, Walia2022EPJC},  Horndeski gravity is defined in terms of the determinant of the four-metric, the Ricci scalar, and the scalar field. By imposing the canonical operation of the scalar field and a finite energy condition, solving the field equations results in a static spacetime solution with spherical symmetry for Horndeski gravity \citep{Esteban2021arXiv, Heydari-Fard2023arXiv}.

\begin{eqnarray}
  ds^2 = -f(r)dt^2 + \frac{dr^2}{f(r)}+r^2\left(d\theta^2 + Sin^2\theta d\phi^2\right),
\label{Horn1}
\end{eqnarray}

\noindent
where $f(r) = 1 - \frac{2M}{r}+\frac{h}{r}ln\left(\frac{2}{2M}\right)$. Here, $M$ is the
mass of the black hole and $h$ is the scalar hair parameter with the dimension of the length.
Horndeski gravity defines the most general scalar-tensor theory with second-order field equations.
The hair parameter $h/M$ can be used to explain different physical results in astrophysical observations
by adding the scalar field to the matrix.

  For the numerical solution of the GRH equations, it is necessary to define the rotating Horndeski black hole and the surrounding spacetime in Boyer-Lindquist coordinates. To achieve this, the Newman-Janis algorithm is used in Eq.\ref{Horn1} \citep{Azreg2014EPJC, Walia2022EPJC}. Thus, the metric of the rotating hairy black hole is given below \citep{Walia2022EPJC}.

\begin{eqnarray}
  ds^2 = -\left(\frac{\bigtriangleup - a^2Sin^2\theta}{\Sigma}\right)dt^2 + \frac{\Sigma}{\bigtriangleup}dr^2
  + \Sigma d\theta^2 + \frac{2a Sin^2\theta}{\Sigma}\left(\bigtriangleup - \left(r^2+a^2\right)\right) dt d\phi \nonumber \\
  + \frac{Sin^2\theta}{\Sigma}
  \left[\left(r^2+a^2\right)^2 -\bigtriangleup a^2 Sin^2\theta \right ] d\phi^2,
\label{Horn2}
\end{eqnarray}

\noindent
where $\bigtriangleup = r^2+a^2-2Mr + h\; r\; ln\left(\frac{r}{2M}\right)$ and
$\Sigma=r^2+a^2 Cos^2\theta$. Here, the values of $h/M$ can have an interval $-2\leq h/M \leq 0$.
The metric can  characterize the rotating black holes for certain values of $h/M$ and $a/M$
discussed in Fig.\ref{hair_param}. The space time around the rotating black hole exhibits a complex dynamics.
When the $h/M \rightarrow 0$, the Horndeski metric given in Eq.\ref{Horn2} goes to the
Kerr black hole solution \citep{Kerr1963PhRvL}.
Furthermore, when only $a/M$ is reduced to zero, it produces a spherically  symmetric hairy
black hole solution. In addition, the metric transitions to the Schwarzschild solution 
when both $h/M$ and $a/M$ are simultaneously set to zero.

In order the solve the GRH equations with Horndeski  space time metric, the lapse function and the shift vectors
should be determined the metric framework. 
The relationship between the four-metric $g_{ab}$ and the three-metric $\gamma_{ij}$,
together with the lapse function $\alpha$ and shift vectors $\beta_i$, is determined using the
following relation\citep{Misner1977}:

\begin{eqnarray}
  \left( {\begin{array}{cc}
   g_{tt} & g_{ti} \\
   g_{it} & \gamma_{ij} \\
  \end{array} } \right)
=  
  \left( {\begin{array}{cc}
   (\beta_k\beta^k - \alpha^2) & \beta_k \\
   \beta_i & \gamma_{ij} \\
  \end{array} } \right),
\label{Horn3}
\end{eqnarray}

\noindent
where the lapse function for Horndeski black hole is

\begin{eqnarray}
  \alpha = \sqrt{-\frac{a^2 - \bigtriangleup}{\Sigma} +
    \frac{a^2\left(-a^2 + \bigtriangleup - r^2\right)^2}{\Sigma^3}}.
\label{Horn4}
\end{eqnarray}

\noindent
The shift vectors can be represented as  

\begin{eqnarray}
  \beta_r = 0, \;\;\;  \beta_{\theta}= 0, \;\;\;  \beta_{\phi}= \frac{a\left(-a^2+\bigtriangleup - r^2\right)}{\Sigma}.
\label{Horn5}
\end{eqnarray}



\section{Initial and Boundary Conditions}
\label{InitBoun}

BHL accretion occurs when gas falls from the
upstream region towards the black hole through the spherical accretion.
As a result, it not only creates an accretion disk around the black hole but also
leads to the formation of physical mechanisms such as shock waves and cones.
These mechanisms can be utilized to comprehend the physical processes underlying
the observed QPOs.
To identify the type of physical mechanisms formed as a result of BHL accretion,
we numerically solve the GRH equations, which describe the behavior of gas within the fixed
spacetime around the  black hole.
In this article, the physical behavior
of the shock cone formed around the rotating Horndeski black hole is examined using the Horndeski spacetime matrix.
The Horndeski black hole reveals the effects of both the rotation of the black hole
and the hair parameters on the formed shock cone. Thus, our aim is to provide explanations
for the physical mechanisms of the observed QPO frequencies, which differ from our
previous studies \citep{Donmez5, Donmez4, Donmez_EGB_Rot, Donmez2023arXiv231013847D, Donmez2024Univ}.

When modeling the shock cone around the Horndeski black hole, the matter is sent from the
upstream region with spherical accretion from the outer boundary of the computational
domain. The physical properties of the sent matter include: $\rho=1$, $C_{\infty} = 0.1$, 
$\Gamma = 4/3$, $V^r= \sqrt{\gamma^{rr}}V_{\infty}cos(\phi)$, $V^{\phi}= -\sqrt{\gamma^{\phi\phi}}V_{\infty}sin(\phi)$
are the rest-mass density, the sound speed, the adiabatic index of the matter, radial velocity,
and angular velocity of the injected matter, respectively. The pressure of the
falling matter is computed using the ideal gas equation of state, $P = (\Gamma - 1)\rho\epsilon$.
In most models, we use the asymptotic speed of $V_{\infty}/c = 0.2$. However, to reveal
the impact of the asymptotic speed on the shock cone and to make comparisons with the
literature, we also employ the  different asymptotic speeds,
as shown in Table \ref{Inital_Con}.
During the formation of the shock cone in BHL accretion,
 $V_{\infty}/c$ plays a significant role in creating an effect. This effect has
been extensively discussed in the literature across the different gravities
\cite{Ruffert1, Foglizzo1, Zanotti1, Donmez6, Donmez5, Koyuncu1, 
  LoraClavijo2, CruzOsorio2020ApJ,Donmez3, Donmezetal2022, CruzOsorio2023JCAP}.
However, in our upcoming article, we will extensively discuss the impact of $V_{\infty}/c$, considering
the scalar hair and the rotation parameter of the black hole as well. At the same time,
we will reveal the effect of the adiabatic index ($\Gamma$) on the formation of shock cones.
Since gas can be more compressible at higher values of $\Gamma$, we will examine the
condition for the forming a bow shock in these calculations.  The formation of the bow shock is
a physical mechanism observed in the case of an ultralight scalar field, under different
$\Gamma$ and $V_{\infty}/c$ \cite{CruzOsorio2023JCAP}.

\begin{table}
\footnotesize
\caption{The initial model adopted for the numerical simulation of Kerr and
  Horndeski metric and some outcomes produced from the numerical results.
  $Model$, $type$, $a/M$, $h/M$, $V_{\infty}/c$, $r_{stag}/M$, $\theta_{sh}/rad$, and $\tau_{ss}$
  are the name of the model,
  the gravity, the black hole rotation parameter, scalar hair parameter in Horndeski metric,
  asymptotic velocity of gas injected from the outer boundary, the position of the stagnation
  point, shock cone opening angle, and time to require to reach the steady-state, respectively.}
 \label{Inital_Con}
\begin{center}
  \begin{tabular}{cccccccc}
    \hline
    \hline

    $Model$        & $type$         & $a/M$ & $h/M$ & $V_{\infty}/c$ & $r_{stag}/M$ & $\theta_{sh}/rad$
    &$\tau_{ss}/M$\\ 
    \hline
    $H04A$         &  $Kerr$        & $0.4$ & $0$    & $0.2$ & $27.2$ & $1.076$ & $2561$\\
    $H04B$         &                & $0.4$ & $-0.5$ & $0.2$ & $14.05$& $0.782$ & $7900$\\
    $H04C$         &  $Horndeski$   & $0.4$ & $-0.8$ & $0.2$ & $10.6$ & $0.561$ & $4600$\\
    $H04D$         &                & $0.4$ & $-1.0$ & $0.2$ & $9$    & $0.344$ & $1900$\\
    $H04E$         &                & $0.4$ & $-1.2$ & $0.2$ & $8.3$  & $NO$    & $NO$ \\
    \hline
    $H06A$         &   $Kerr$       & $0.6$ & $0$    & $0.2$ & $27$   & $1.052$ & $2400$\\
    $H06B$         &                & $0.6$ & $-0.1$ & $0.2$ & $23.3$ & $1.028$ & $1900$\\
    $H06C$         &  $Horndeski$   & $0.6$ & $-0.2$ & $0.2$ & $20.2$ & $0.979$ & $1700$\\
    $H06D$         &                & $0.6$ & $-0.4$ & $0.2$ & $15.9$ & $0.88$  & $1400$\\
    $H06E$         &                & $0.6$ & $-0.6$ & $0.2$ & $12.5$ & $0.708$ & $4550$\\
    $H06F$         &                & $0.6$ & $-0.8$ & $0.2$ & $10$   & $0.537$ & $4150$\\
    $H06V_{\infty}04$&                & $0.6$ & $-0.8$ & $0.4$ & $7.1$  & $0.684$ & $4000$ \\
    \hline
    $H09A$         &   $Kerr$       & $0.9$ & $0$    & $0.2$ & $26.85$& $1.052$ & $2350$\\
    $H09B$         &                & $0.9$ & $-0.1$ & $0.2$ & $23.3$ & $1.027$ & $2300$\\
    $H09C$         &  $Horndeski$   & $0.9$ & $-0.15$& $0.2$ & $21.8$ & $1.003$ & $2100$\\
    $H09D$         &                & $0.9$ & $-0.2$ & $0.2$ & $20.5$ & $0.979$ & $1900$\\
    $H09V_{\infty}02$&                & $0.9$ & $-0.25$& $0.2$ & $19.35$& $0.954$ & $1800$\\    
    $H09E$         &                & $0.9$ & $-0.27$& $0.2$ & $18.5$ & $0.93$  & $1700$\\
    \hline
    $H09V_{\infty}01$&                & $0.9$ & $-0.25$& $0.1$ & $27.3$ & $1.077$ & $2080$ \\
    $H09V_{\infty}04$&                & $0.9$ & $-0.25$& $0.4$ & $9.4$  & $0.806$ & $970$ \\
    $H09V_{\infty}06$&                & $0.9$ & $-0.25$& $0.6$ & $5.8$  & $0.757$ & $370$ \\        
    \hline
    \hline
  \end{tabular}
\end{center}
\end{table}

\begin{figure*}
  \center
    \psfig{file=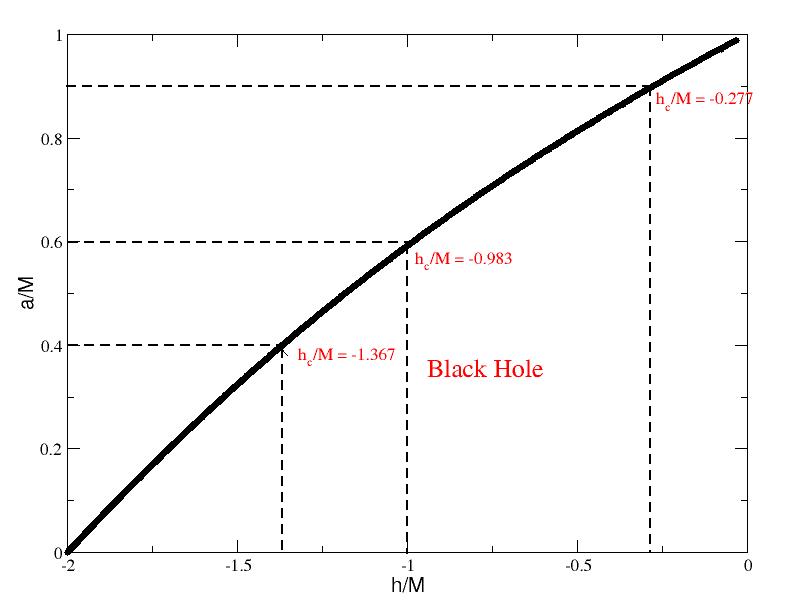,width=14.0cm,height=13.0cm} 
    \caption{The critical values of the scalar hair parameter ($h/M$) that the Horndeski matrix depends on,
      according to the black hole rotation parameter ($a/M$). 
      There exists a maximum possible value for $h/M$ for every given value of $a/M$.
      Black hole solutions exist for values of $a/M$ and $h/M$  that have been located under the curve
      shown in the figure. The values in Table \ref{Inital_Con} are selected according to this plot.}
\label{hair_param}
\end{figure*}

In this study, the scalar hair parameter ($h/M$) in Horndeski gravity constitutes the main theme
of our work. We will investigate the effect of this parameter on the mass accretion rate and
the creation of the shock cone around the rotating black holes. However, not every value of $h/M$
in Horndeski gravity results in a rotating black hole. There exists a critical value of
$h/M$ for each rotation parameter. Only if $h/M <h_{c}/M$  is there a black hole solution.
Fig.\ref{hair_param} shows the $h_{c}/M$ for  different rotation parameters. $h_{c}$ are determined from
the analytic calculation using the spacetime metric of Horndeski.
For every $h/M$ value that lies below the curve shown in Fig.\ref{hair_param}, the black hole is formed.
The $h/M$ values given in Table \ref{Inital_Con}  have been selected according to this situation.

The computational domain used to construct a stable shock cone around the Horndeski
black hole is $ r\in [2.3M, 100M]$ and $\phi \in [0, 2\pi]$. 
A grid spacing of equal distance in both radial and angular directions are utilized.
In the radial direction, $1024$ points are used, while in the angular direction,
this value is $256$.
For all models, a maximum time of $t_{max}=35000M$ has been set. The reason for running the codes for
such a long duration is to examine the behavior of the shock cone after it has reached the
steady-state. This is crucial because  QPO frequencies are determined based on this behavior.

As seen in Table \ref{Inital_Con}, the shock cone reaches a steady-state
around $t=2000M$. Subsequently, it has exhibited instability behavior around the
certain value. Drawing from extensive experience with QPOs, we are confident that QPO frequencies do not
depend on grid resolution \citep{Donmez3, Donmez_EGB_Rot, Donmezetal2022, Donmez2024Univ}.

At the outer boundary of the computational domain in radial direction, two different
situations are present. Since gas is continuously injected into the computational domain
from the upstream region, the same value has been used in the  boundaries of the upstream
region. However, an outflow boundary condition has been applied to expel the numerical issues
at the boundary caused by the gas reaching the downstream region.
On the other hand, an outflow boundary condition has been implemented at the inner boundary
of the computational domain. This means that the gas approaching the horizon is expelled
from the computational domain, signifying the matter falling into the black hole.
Periodic boundary conditions are enforced in the $\phi$-direction to preserve the symmetry of
the physical solutions..


\section{Results}
\label{Result1}

In this section, we model the dynamics of the shock cone numerically and reveal its formation around
the rotating Horndeski black hole as a result of BHL accretion.
We also attempt to understand the dynamic behavior of the shock cone on the downstream side of the
computational domain.
We uncover how the dynamic structure of the shock cone and the physical
behavior of the trapped and excited  QPOs  change with the black hole
rotation parameter ($a/M$), scalar hair parameter of the Horndeski gravity ($h/M$),
and asymptotic velocity of the injected matter injected from the outer boundary
in the upstream region ($V_{\infty}/c$).

  BHL accretion occurs when matter falls onto an object moving supersonically through a uniform medium, leading to non-spherical accretion of matter only on one side of the object. On the other side of the object, matter accretion is very weak. This is the fundamental difference that distinguishes BHL accretion from spherical Bondi accretion. Numerical calculations conducted over many years have shown that BHL accretion results in the formation of a shock cone on the opposite side of the object \citep{Edgar1, Foglizzo1, Koyuncu1, Donmez5, LoraClavijo2}. This cone itself demonstrates that the accretion is non-spherical \citep{Donmez6, CruzOsorio2020ApJ}. If the object is a black hole, the shock cone formed by this accretion can extend to the inner boundary of the computational domain, i.e., close to the black hole's horizon, making the shock cone an important physical mechanism for revealing the frame-dragging effect in strong gravitational fields. Additionally, the oscillation modes trapped within the cone are an important physical mechanism for producing QPO frequencies. Our current studies and existing literature show that BHL accretion is a significant mechanism because of these features. Shock cones that can form very close to the black hole's horizon in strong gravitational fields emerge as an important mechanism for explaining many observational data \citep{Donmez6, Zanotti1, Wenrui1, CruzOsorio2023JCAP}. Furthermore, as demonstrated in this study, it is also an important mechanism for revealing the effect of scalar fields \citep{Donmez2024Submitted} and for testing alternative gravities \citep{Donmez3, Donmezetal2022, Donmez2024Univ} on the physical mechanisms around the black hole. Thus, the differences between modified gravities and Kerr gravity can be understood through the BHL accretion mechanism.

\subsection{Numerical Results }
\label{Result2}

\begin{figure*}
  \center
    \psfig{file=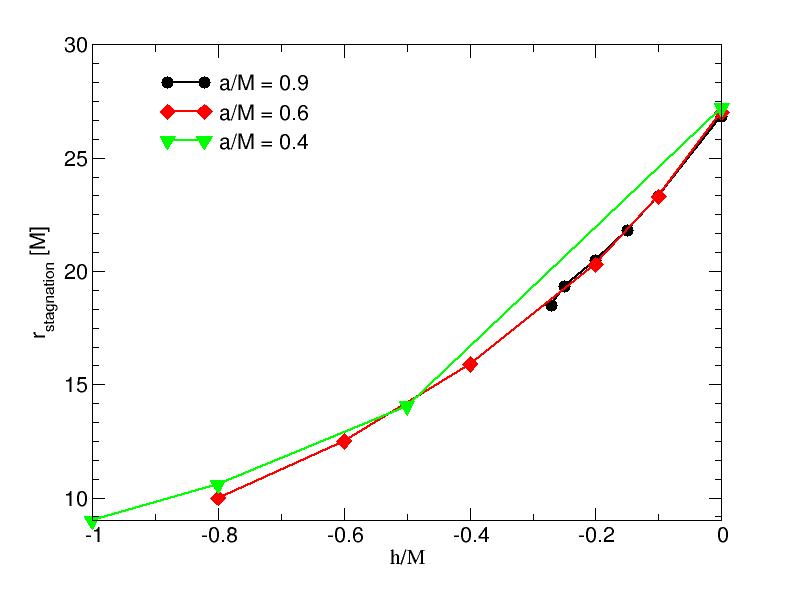,width=14.0cm,height=12.0cm} 
    \caption{The radial position of the stagnation point as a function of Horndeski  hair parameter.
      The changes in the stagnation point under different black hole rotation parameters and possible
      hair parameters have been demonstrated.}
\label{stagnation_all}
\end{figure*}

Understanding  the behavior of the stagnation point around the black hole helps determine
the accretion speed of the matter falling towards the black hole and thus aids in understanding
the spectral energy distribution. Fig.\ref{stagnation_all} shows the behavior of the stagnation
point for different black hole spin parameters according to the scalar hair parameter. As seen
in the figure, while the behavior of the stagnation point undergoes significant changes with the
variation of the hair parameter, this change is not  obvious in the case of different
spin parameters. As the hair parameter increases in the negative direction, as seen in
Fig.\ref{stagnation_all}, the stagnation point approaches the black hole  horizon with an
exponential decrease. This change, as discussed in more detail later, leads to the sweeping outwards
of the shock cone, thereby causing a change or complete disappearance of the QPO frequencies.

On the other hand, as seen in Table \ref{Inital_Con}, an increase in the asymptotic speed
also causes the stagnation point to approach the black hole horizon. However, the behavior of the shock
cone in this case is completely different from that in the case of the hair parameter. In this
scenario, more matter is accumulated within the shock cone, especially close to the event horizon
leading to a change and then disappearance of the QPO frequencies.

\begin{figure*}
  \center
    \psfig{file=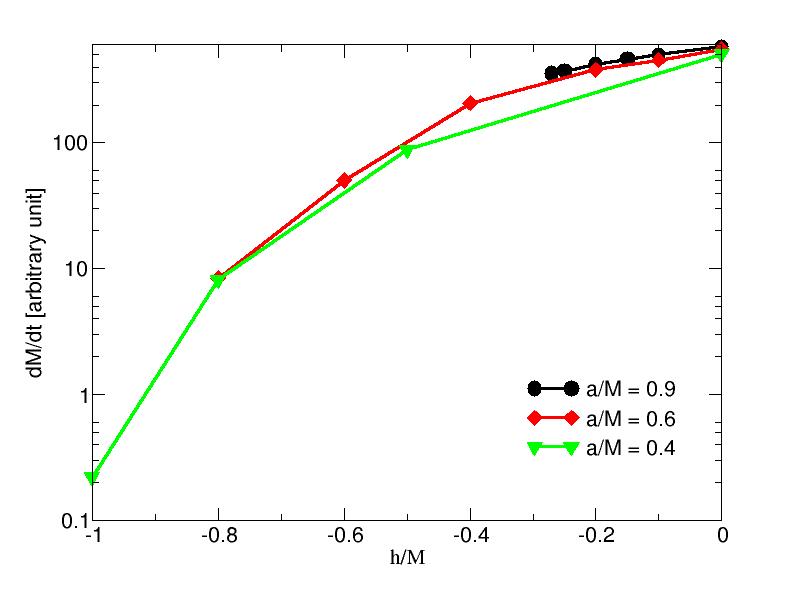,width=14.0cm,height=12.0cm} 
    \caption{The change in the mass accretion rate depending on Horndeski hair parameters
      for different black hole rotation parameters with $V_{\infty}/c=0.2$.
      Each point on the figure has been calculated long after the shock cone reaches to the steady-state
      at $r=4M$.}
\label{MA_tmax_all}
\end{figure*}

The variation in the maximum value of the mass accretion rate after
the shock cone has reached a steady-state is presented as a function of $h/M$ In Fig.\ref{MA_tmax_all}.
In the numerical calculations, the mass accretion is  computed close to the black hole horizon,
specifically at $r/M=4M$. Subsequently, the changes with the scalar hair and the black hole
rotation parameter are depicted in Fig.\ref{MA_tmax_all}. Again, the impact of the rotation
parameter on the mass accretion rate has not been observed in a pronounced manner. However, the scalar
hair parameter has significantly reduced the mass accretion rate. Figs.\ref{stagnation_all} and
\ref{MA_tmax_all} show a complete correlation between the behaviors of the stagnation point
and the mass accretion rate, which is an expected outcome.
The most fundamental difference between them is that the stagnation point creates an
inwardly curved line, while the mass accretion rate forms an outwardly curved line.

Understanding the mass accretion rate  around the black holes significantly contributes to
explaining many astrophysical phenomena. For instance, a high mass accretion rate leads to
matter falling into the black hole and, as a result, causes electromagnetic emissions in the
strong gravitational field. Additionally, as matter falls into the black hole regularly, it
increases the mass of the black hole. Thus, this can explain the existence of
massive black holes.

\subsection{The case of $a/M = 0.6$ }
\label{casea06}

\begin{figure*}
  \vspace{1cm}
  \center
     \psfig{file=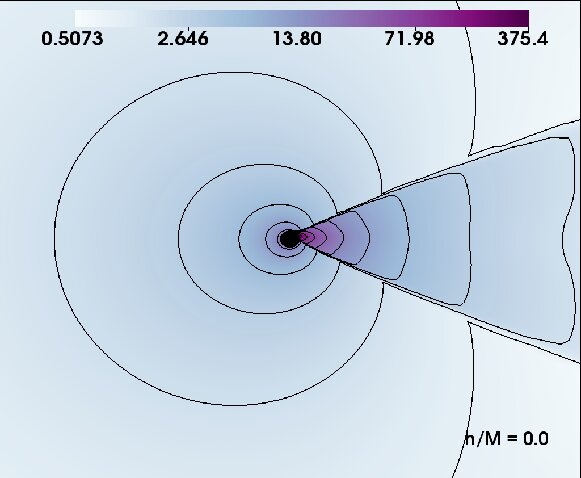,width=7.5cm}
    \psfig{file=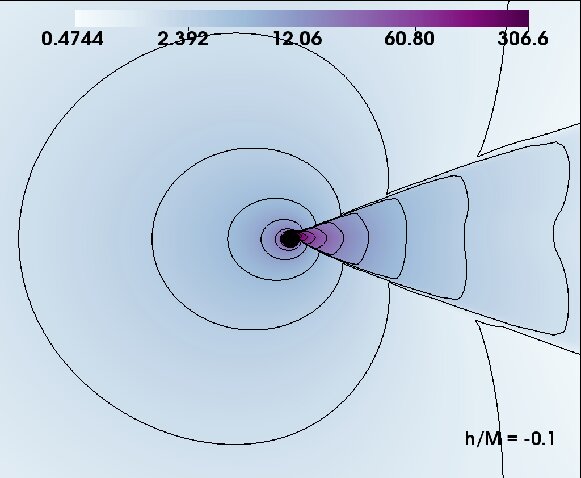,width=7.5cm}
    \psfig{file=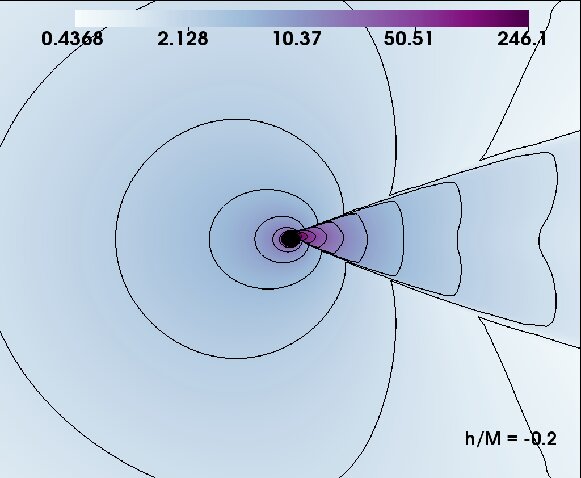,width=7.5cm}
    \psfig{file=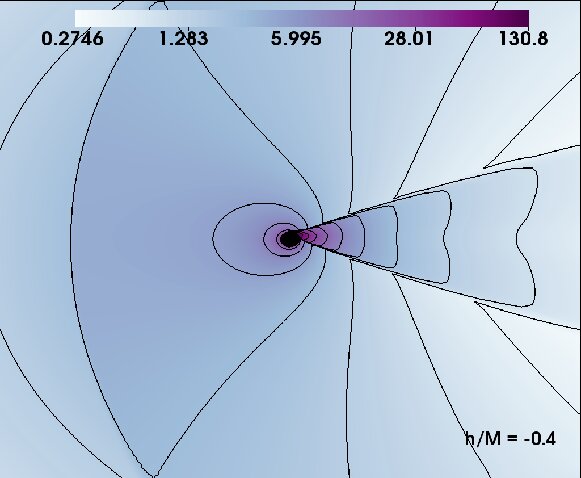,width=7.5cm} 
    \psfig{file=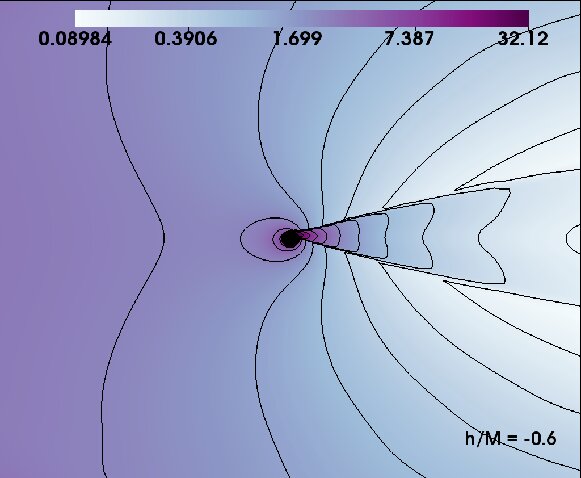,width=7.5cm}
    \psfig{file=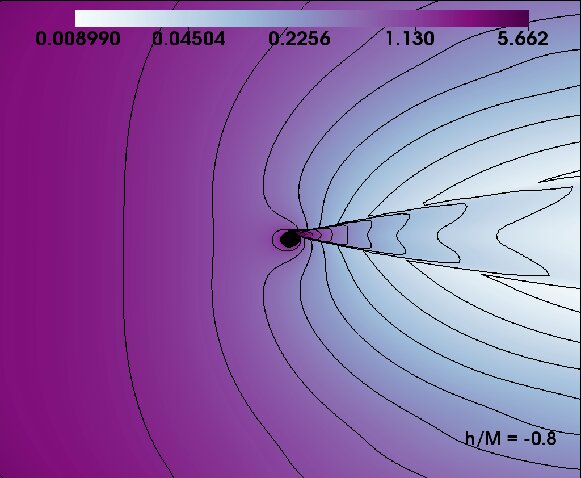,width=7.5cm}
    \caption{The rest-mass density variations across the different Horndeski parameters
      on the equatorial plane for $a/M = 0.6$ are shown, using the both color and contour lines
      at the end of the numerical evolution $t=35000M$. As seen in Table \ref{Inital_Con},
      each model has reached the steady-state well before the maximum time.
      In particular, to observe the dynamic structure of the shock cone close to the black hole in more details,
      the minimum and maximum limits on the $x$ and $y$ axes have been set from $-70M$ to $70M$.}
\vspace{1cm}
\label{a06_color}
\end{figure*}

The morphology is shown in Fig.\ref{a06_color} long after the shock cone has reached
the steady-state ($t \sim 2000M$), and at a much later time ($t=32000M$). The figure illustrates
the change in the rest-mass density  and the formation of the shock cone around the Kerr and
Horndeski black holes with a rotation parameter of $a/M=0.6$, using a colormap and isocontours.
The top left graph displays the  morphology of the rest-mass density  in Kerr gravity, while the others showcase the
dynamics of the shock cone around the Horndeski black hole under different scalar hair parameters
($h/M > h_c/M=-0.983$ ).

The BHL accretion mechanism has been extensively studied in
the literature across various gravitational models, such as Kerr, Einstein-Gauss-Bonnet,
and Hartle-Thorne \cite{Donmez5, Zanotti1, Donmez6, Donmez3, Donmez_EGB_Rot,
  CruzOsorio2023JCAP, Donmez20}. It has been observed that the shock cone forms regardless
of the gravity model employed. BHL accretion results from the matter falling towards the black hole
supersonically from the upstream region, leading to the formation of the shock cone on the
downstream side. The shock cone has been observed at various values of $h/M$ as seen in
Fig.\ref{a06_color}.
Although the shock cone forms for every value of $h/M$, it has been observed that, unlike with
other gravity models, the scalar hair parameter in Horndeski gravity significantly affects the dynamic structure of the shock cone. 
When considering aspects such as the cone opening angle
and the stagnation point, as shown in Table \ref{Inital_Con}, it is evident
in Fig.\ref{a06_color} that as $h/M$ increases in the negative direction,
the dynamic structure of the shock cone  undergoes a significant change.

The interaction of the scalar field with spacetime alters the behavior of matter on the
downstream side. As $h/M$ increases in the negative direction,
indicating an intensification of the scalar field, 
the opening angle of the formed shock cone decreases.
Concurrently, the rate at which trapped matter falls into the black hole diminishes with $h/M$.
In essence, as the force created by the scalar field intensifies with $h/M$ becoming more negative,
the matter begins to move further away from the black hole. This shift also causes the stagnation
point to move closer to the black hole horizon. These physical changes are clearly observable
in the  $h/M =-0.6$ and $h/M =-0.8$ models. This intriguing result suggests that the shock cone
is on a trajectory towards disappearance as $h/M$ varies.
Consequently, the change in the dynamic structure
 of the shock cone has resulted in alterations to the oscillation frequencies of trapped
 pressure-based and radial-based QPO modes, and even their disappearance.
The possible QPOs are discussed in detail in Section \ref{Possible_QPO}.

\begin{figure*}
  \center
    \psfig{file=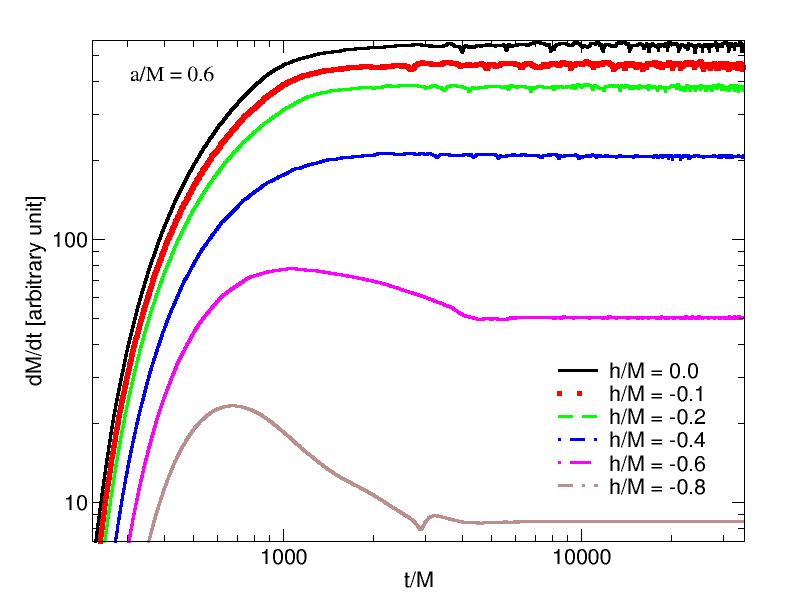,width=14.0cm,height=13.0cm} 
    \caption{The change in mass accretion rate over time for $a/M=0.6$, demonstrated for
      each Horndeski hair parameter. The effect of the scalar hair parameter on the dynamics of the
      mass accretion rate around the black hole is shown.}
\label{MA_a06}
\end{figure*}

The variation in the mass accretion rate near the black hole horizon, at $r=4M$,
is depicted in Fig.\ref{MA_a06}. It is evident that the shock cone has reached
the steady-state in all models. However, the time required to reach this steady-state
varies, as indicated in Table \ref{Inital_Con}.
The behavior of the mass accretion rate
confirms the discussions presented in Fig.\ref{a06_color} and the numerical results listed
in Table \ref{Inital_Con}.
The influence of the scalar hair parameter, which leads to
a decrease in the amount of matter falling towards the black hole, is distinctly
visible in Fig.\ref{MA_a06}. As $h/M$ becomes more negative, the stagnation point shifts closer
to the black hole, thereby deflecting more matter away from it.
Consequently, there is a reduction in the mass accretion rate, implying that the quantity of matter being
absorbed by the black hole diminishes over time. Notably, as demonstrated in Fig.\ref{MA_a06}
the dynamic structure of the shock cone at $h/M=-0.6$
and $h/M=-0.8$ exhibits significant variances when compared to other values of  $h/M$.
This suggests that the behavior of the shock cone undergoes  more rapid changes once  the
stagnation point descends a certain threshold, which occurs for $h/M<-0.4$.
Moreover, post-reaching the steady state, the shock cone displays instability, and the mass accretion
rate hints at the emergence of QPOs. However, as discussed in Section \ref{Possible_QPO}, no QPO
frequencies are observed in the model with $h/M=-0.8$.

\subsection{The case of $a/M = 0.4$ }
\label{casea04}

\begin{figure*}
  \vspace{1cm}
  \center
     \psfig{file=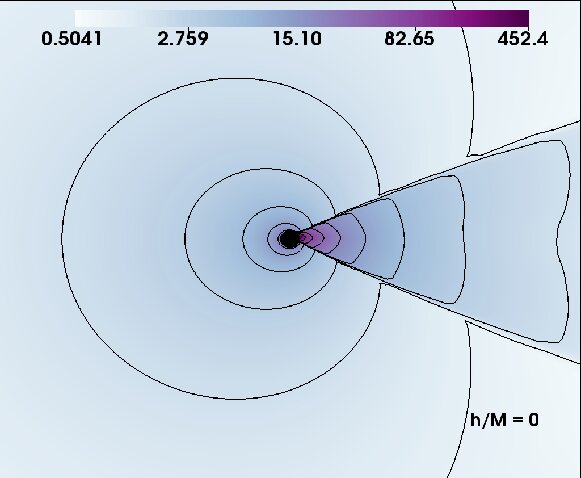,width=7.5cm}
    \psfig{file=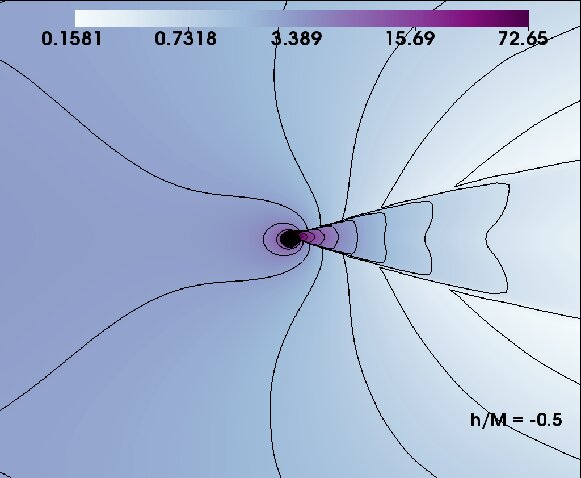,width=7.5cm}
    \psfig{file=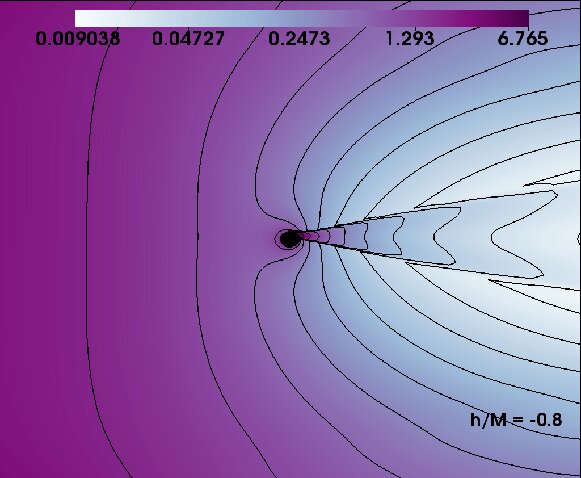,width=7.5cm}
    \psfig{file=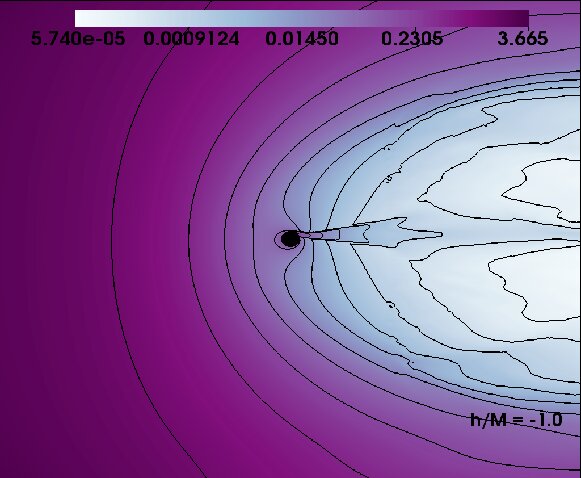,width=7.5cm} 
    \caption{Same as in Fig.\ref{a06_color} but for the $a/M=0.4$.}
\vspace{1cm}
\label{a04_color}
\end{figure*}

To uncover the structure of the shock cone formed around the Horndeski black hole due to the interaction of
spacetime with a strong scalar field, we model the scenario with a small rotation parameter. For $a/M = 0.4$,
Fig. \ref{a04_color} demonstrates how the dynamic structure of the shock cone changes in the presence of both
weak and strong scalar fields, while also comparing it to the Kerr solution.

The growth of the scalar field parameter $h/M$ in the negative direction has significantly altered the structure
of the shock cone formed on the downstream side of the computational domain around the black hole. Due to the
increased effect of the field as $h/M$ changes, more matter within the shock cone has been pushed away from the
black hole. As a result, the shock cone has entered a process of disappearance. In Section \ref{special_case},
the stronger scalar field defined with $h/M=-1.2$ has been modeled, and it is observed that the cone completely
vanishes. Consequently, QPO frequencies, which could have been generated by the pressure and radial modes trapped
entirely within the shock cone, do not form. Indeed, our results for the rotation parameters $a/M = 0.4$ and
$a/M = 0.6$ strongly support the notion that the shock cone is one of the most significant physical mechanisms
that can be proposed to explain observational QPOs.

\subsection{A special Case: $a/M = 0.4$ and $h/M=-1.2$}
\label{special_case}

\begin{figure*}
  \vspace{1cm}
  \center
     \psfig{file=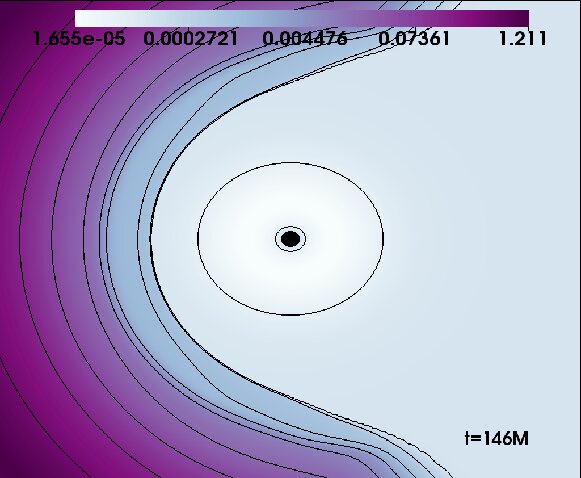,width=7.5cm}
    \psfig{file=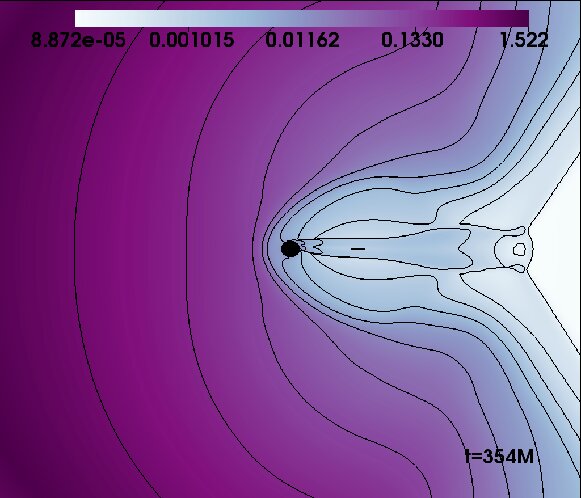,width=7.5cm}
    \psfig{file=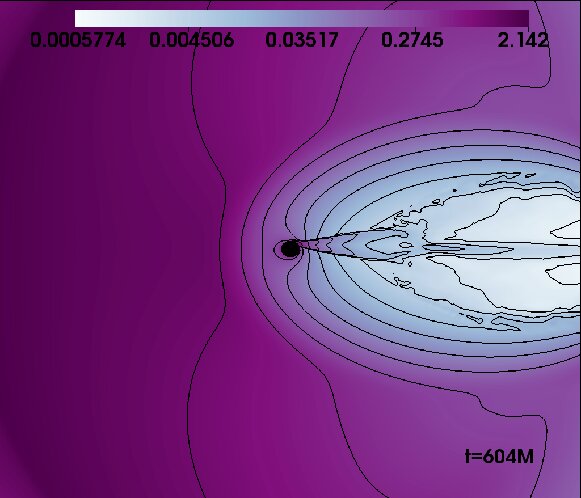,width=7.5cm}
    \psfig{file=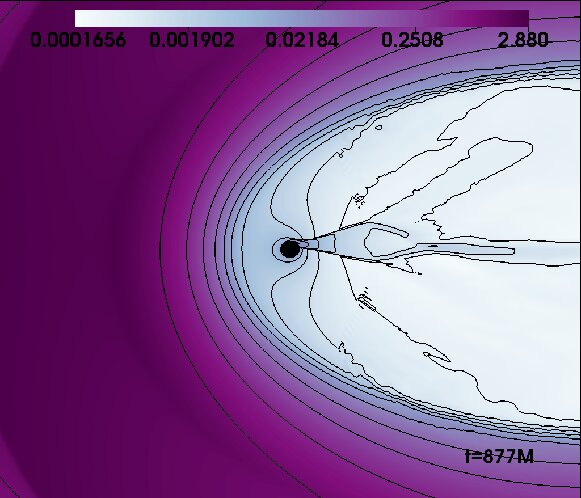,width=7.5cm} 
    \psfig{file=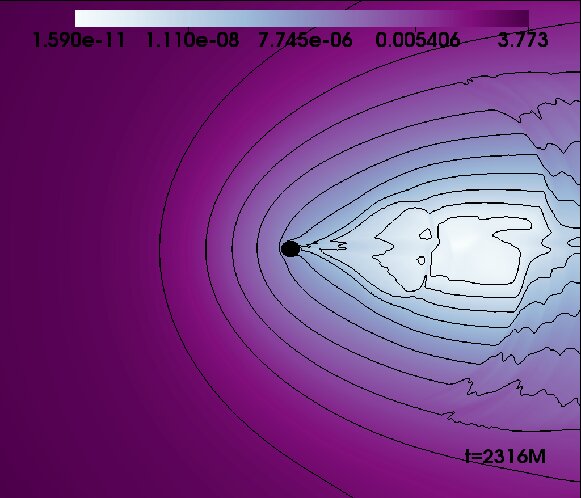,width=7.5cm}
    \psfig{file=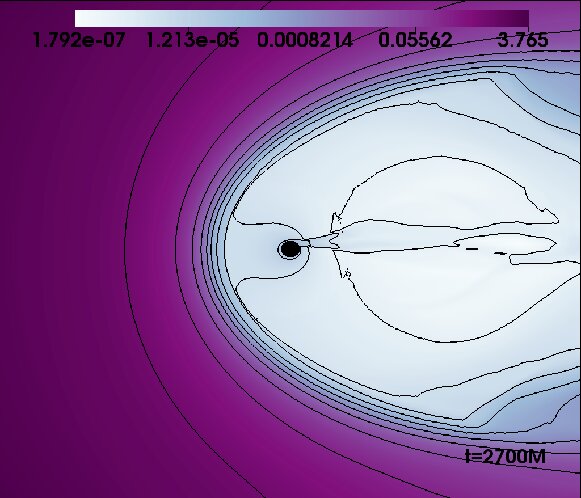,width=7.5cm}    
    \caption{The Color and counter maps  of the rest-mass density in 2D for $a/M=0.4$ with $h/M=-1.2$.
      Each snapshot shows the change in the shock cone  dynamic structure at different times of the same model.
      It demonstrates how the shock cone on the downstream side disappears due to the stagnation point
      getting closer to  the black hole. At the same time, it shows that the conditions in the middle
      and bottom rows exhibit similar behavior. Namely, in the region where the shock cone is located,
      an oscillation state encountered, for the first time, has been observed.
    }
\vspace{1cm}
\label{a04h12_1}
\end{figure*}

\begin{figure*}
  \center
    \psfig{file=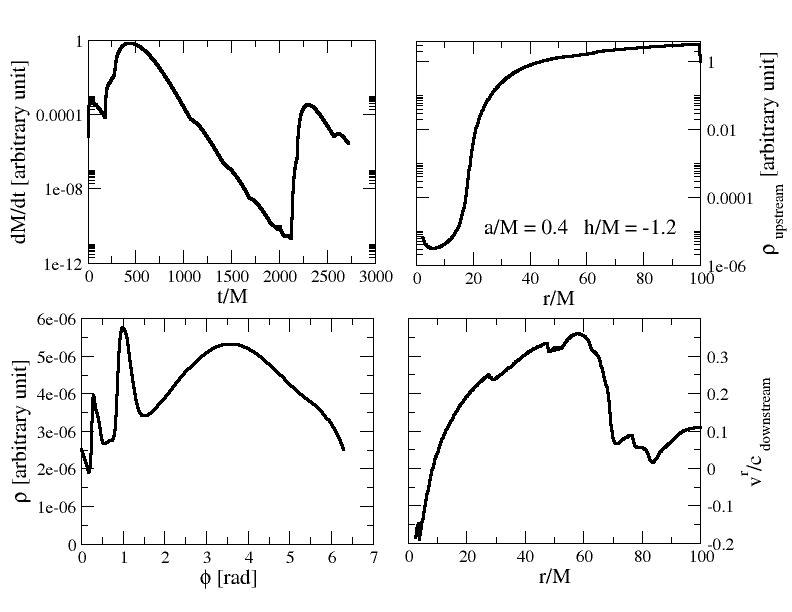,width=14.0cm,height=13.0cm} 
    \caption{ The variation of the different values of the shock cone in the
      vicinity of the Horndeski black hole for the $a/M=0.4$ with $h/M=-1.2$.
      Upper left: This section illustrates the changes in the mass accretion rate.
      Upper right: Here, the rest-mass density on the upstream side of the computational
      domain is depicted along the radial direction at $\phi = 3.14$ rad. Lower left:
      A one-dimensional representation of the rest-mass density is shown along the azimuthal
      direction, positioned at $r=2.68M$, very near the black hole horizon. Lower
      right: Displayed is the radial velocity along the radial axis
      at $\phi =0.024$ rad in downstream side.}
\label{a04h12_2}
\end{figure*}

The possible analytical values of the scalar hair parameter are presented in Fig.\ref{hair_param}. As the
rotation parameter of the black hole increases, $h/M$ approaches zero. This implies that models of rapidly
rotating black holes have a narrower range of $h/M$ compared to other rotation cases. For the slowly rotating
black hole model, specifically $a/M=0.4$ with $0\geq h/M > h_c/M =-1.3670$, it is beneficial to examine the
dynamical changes of the shock cone with respect to $h/M$ during BHL accretion. Therefore, we model the extreme
cases of the scalar field with $h/M=-1.2$. In Fig.\ref{a04h12_1}, the formation of the shock cone at $h/M=-1.2$
is depicted. This plot illustrates the cone formation over time. However, due to the strong interaction of the
scalar field with spacetime, it is observed that all matter reaching the downstream side of the computational
domain is pushed towards the outer boundary of the computational region, away from the black hole. Consequently,
in the case of $a/M=0.4$ with $h/M=-1.2$, no shock cone forms, and all matter reaching the downstream region is
observed to move away from the black hole. Due to the chaotic behavior of the dynamics, the code crashed after a
certain period. Fig.\ref{a04h12_2} again displays the behaviors of different physical parameters for the same
model in one dimension. As can be seen, no shock cone has formed, indicating that no mechanism capable of
trapping QPO modes has been established.

Sections \ref{M87} and \ref{Horndeski_Mass_QPO} have already discussed the possible $h/M$ values for
$M87^*$ \cite{Afrin2022ApJ} and GRS 1915+105 \cite{Rawat2019ApJ, Misra2020ApJ} sources. The observational
results obtained from these sources have been compared with theoretical outcomes to define the possible range
of $h/M$. Considering the possible $h/M$ values for these sources, it is observed that $h/M=-1.2$
significantly deviates from the values calculated based on observations. In fact, the calculations demonstrate
the accuracy of the numerical model results. Since no shock cone forms at $h/M=-1.2$, it is not possible
to form QPOs, and consequently, reliable observation in this model is unattainable.

\subsection{The case of $a/M = 0.9$ }
\label{casea09}

\begin{figure*}
  \vspace{1cm}
  \center
     \psfig{file=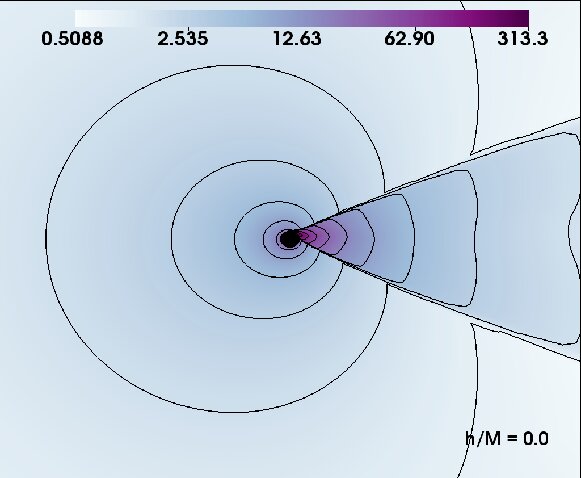,width=7.5cm}
     \psfig{file=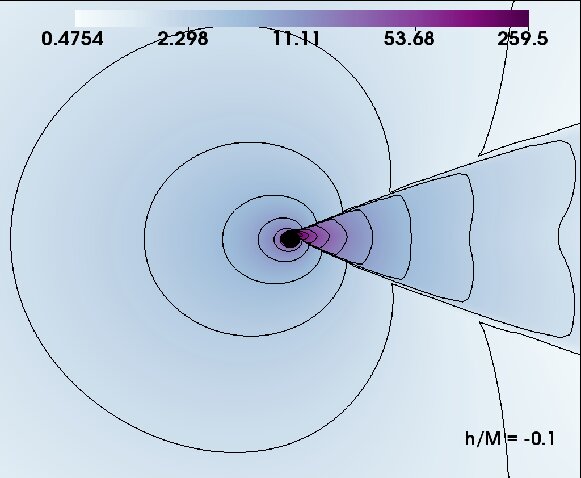,width=7.5cm}
    \psfig{file=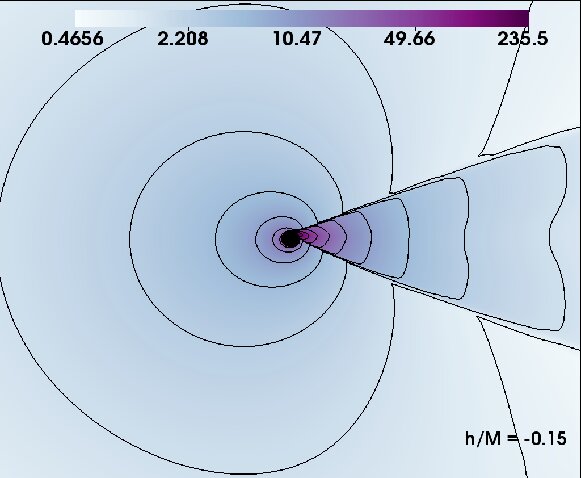,width=7.5cm}
    \psfig{file=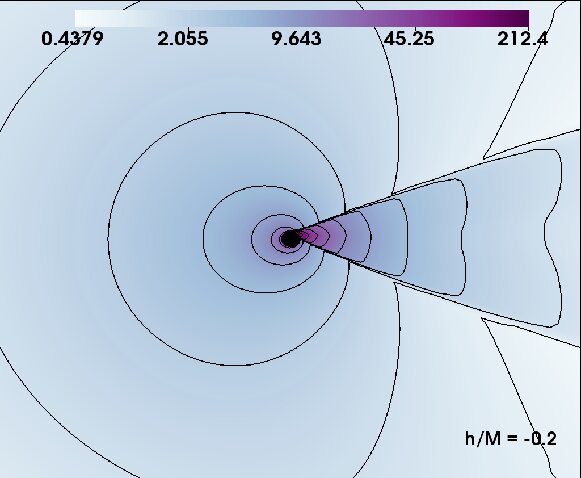,width=7.5cm}
    \psfig{file=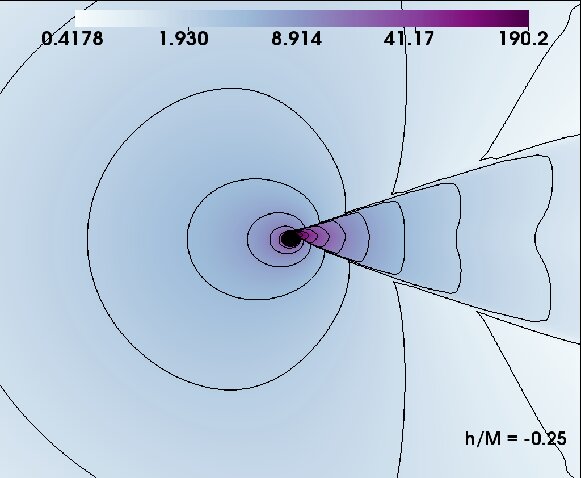,width=7.5cm}
    \psfig{file=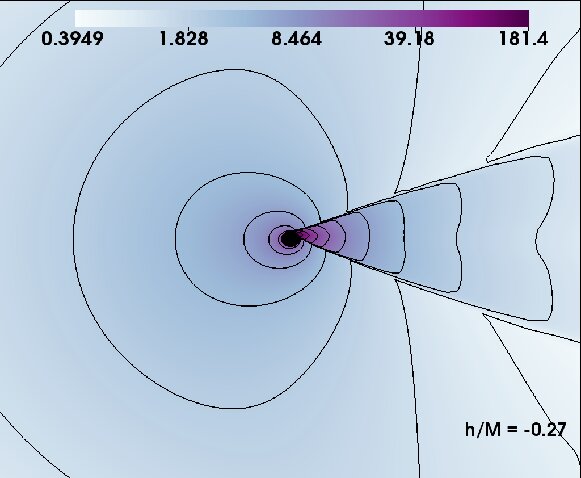,width=7.5cm}    
    \caption{Same as in Fig.\ref{a06_color} but for the $a/M=0.9$. As seen in Fig.\ref{hair_param},
      due to the limited value of $h/M\leq-0.277$, the maximum value of the hair parameter used
      for the case is $-0.27$.}
\vspace{1cm}
\label{a09_color}
\end{figure*}

In Fig.\ref{a09_color}, we model the morphology of the shock cone around a rapidly rotating ($a/M=0.9$)
Horndeski black hole with scalar hair. As observed in Fig.\ref{a09_color}, given that the critical value of
the hair parameter for this rotation parameter is $h_c/M=-0.27$, we explore the physical structure of the shock cone
at various $h/M$ values. In this scenario, since the scalar field cannot be as strong as it is for other rotation
parameters, the structure of the shock cone is maintained across all models. However, we noted a tendency for
the maximum density of the cone to decrease as $h/M$ becomes more negative. This alteration impacts the formation of
pressure and radial modes, leading to changes in the intensity and frequencies of the excited modes. Specifically,
the frequencies increase as $h/M$ decreases. The variation of QPO frequencies for different $h/M$ values in
the case of $a/M=0.9$ is discussed in detail in Section \ref{Possible_QPO}.

\begin{figure*}
  \center
    \psfig{file=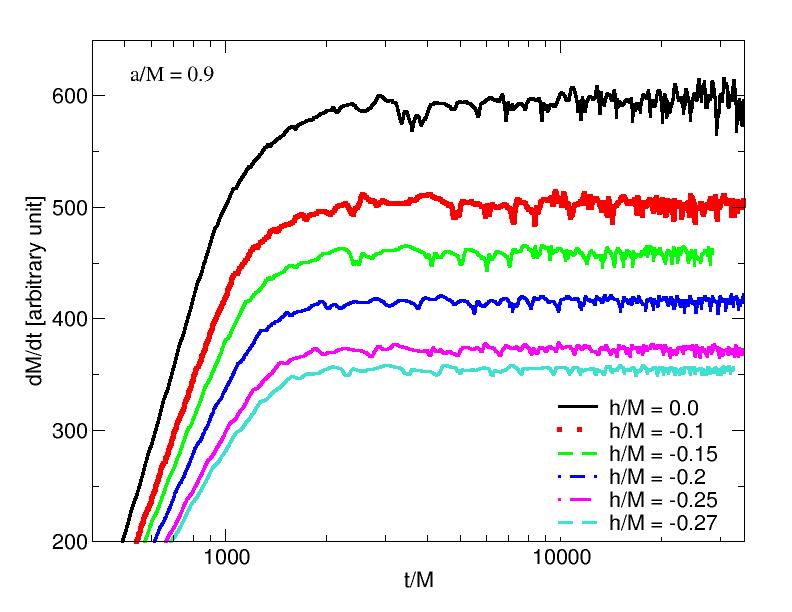,width=14.0cm,height=13.0cm} 
    \caption{Same as in Fig.\ref{MA_a06} but it is for $a/M=0.9$}
\label{MA_a09}
\end{figure*}

As seen in the case of $a/M=0.9$, the variation in the mass accretion rate for different $h/M$ values
is depicted in Fig.\ref{MA_a09}. As $h/M$ becomes more negative, indicating an increase in the intensity
of the scalar field, more matter is pushed outwards in the region where the shock cone is located,
leading to a decrease in the mass accretion rate. This clearly signals a change in the physical structure
of the shock cone. Naturally, this change also impacts the QPO frequencies.

\subsection{The case of $a/M = 0.9$ with $h/M=-0.25$ for Different $V_{\infty}/c$}
\label{a09h025_diff_V}

\begin{figure*}
  \vspace{1cm}
  \center
     \psfig{file=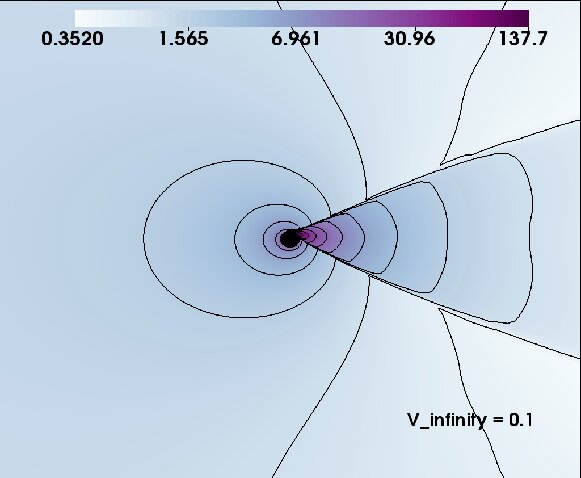,width=7.5cm}
     \psfig{file=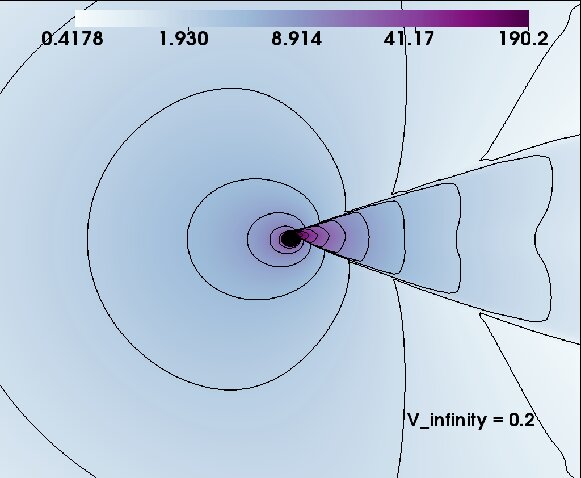,width=7.5cm}
      \vspace*{0.9cm} 
    \psfig{file=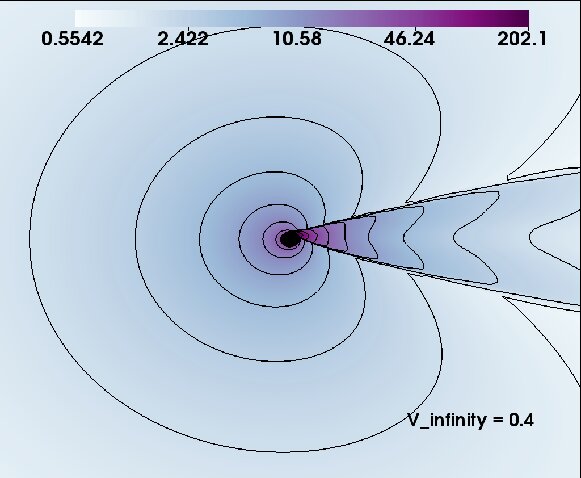,width=7.5cm}
    \psfig{file=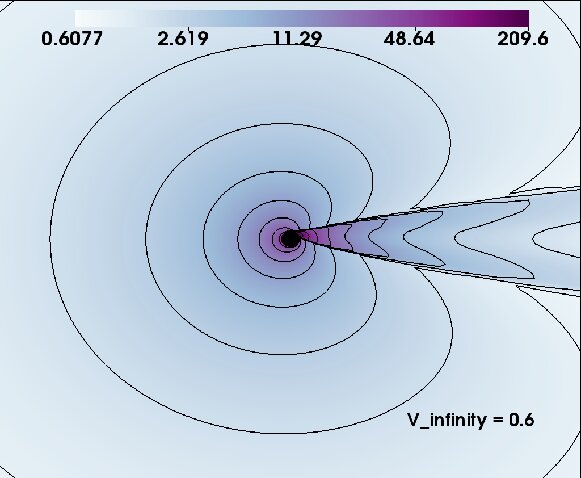,width=7.5cm}
    \caption{Same as in Fig.\ref{a09_color} but for the $a/M=0.9$ with $h/M=-0.25$
      for various values of asymptotic velocity $V_{\infty}/c$ to show the effect of the velocity of
      the matter injected from outer boundary to the shock cone dynamics.
    Each snapshot has been plotted at the maximum time of the simulation.}
\vspace{1cm}
\label{a09_color_Diff_v}
\end{figure*}

\begin{figure*}
  \center
    \psfig{file=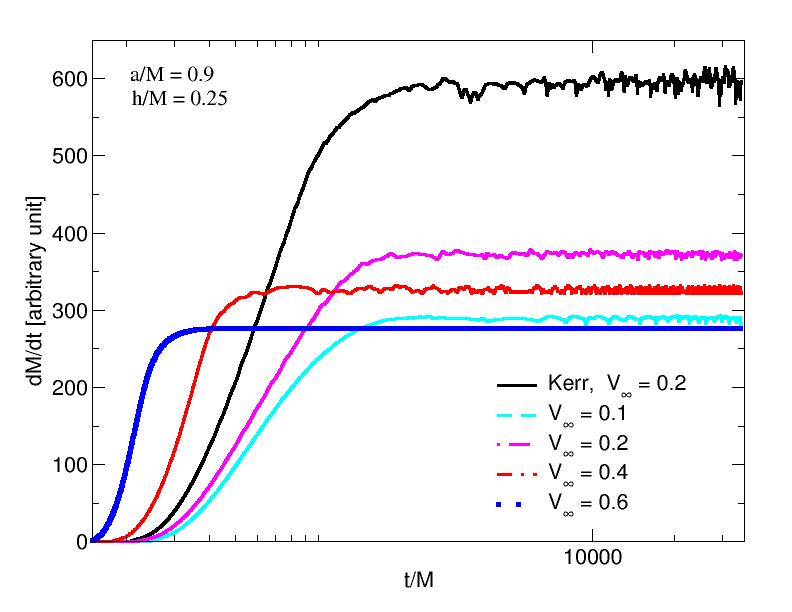,width=14.0cm,height=13.0cm} 
    \caption{Same as in Fig.\ref{MA_a06} but it is for $a/M=0.9$ with $h/M=-0.25$
    for various values of asymptotic velocity $V_{\infty}/c$.}
\label{MA_a09_Diff_v}
\end{figure*}

It is known from studies on Kerr \cite{Ruffert1, Zanotti1, Koyuncu1, LoraClavijo2} and
EGB \cite{Donmez3, Donmez_EGB_Rot} gravity models that the asymptotic speed of matter falling into the black hole
supersonically during BHL accretion significantly impacts the mass accretion and the physical structure of the
shock cone. In this study, for the rapidly spinning black hole model with $a/M=0.9$ and a hair parameter close to
the critical value of $h/M=-0.25$, we demonstrate how the mass accretion (as shown in Fig.\ref{MA_a09_Diff_v})
and the dynamics of the shock cone (as shown in Fig.\ref{a09_color_Diff_v}) vary depending on the asymptotic speed.

As shown in Fig.\ref{a09_color_Diff_v} and Table \ref{Inital_Con}, at the lowest value of the asymptotic speed,
both the shock cone opening angle and the time required to reach the steady state are large. However, these values
decrease as the asymptotic speed increases. Concurrently, the stagnation point moves closer to the black hole horizon,
while the rest-mass density of the matter trapped within the shock cone increases. At $V_{\infty}/c=0.1$ and
$V_{\infty}/c=0.2$, the dynamics of the shock cone and the QPO modes excited within it exhibit behavior similar to
that observed with moderate $h/M$ values in the $a/M=0.4$ and $a/M=0.6$ models. Conversely, in the $V_{\infty}/c=0.4$
and $V_{\infty}/c=0.6$ models, the oscillation modes of the shock cone are notably distinct, as discussed in detail
in Section \ref{Possible_QPO}. Specifically, in the case of $V_{\infty}/c=0.6$, the decrease in the cone opening
angle and the stagnation point closer proximity to the black hole horizon lead to the complete disappearance of
oscillations in the mass accretion rate, as seen in Fig.\ref{MA_a09_Diff_v}. As a result, no QPO mode is excited
for $V_{\infty}/c=0.6$. Interestingly, in the case of $V_{\infty}/c=0.4$, only the $f_{sh}$ mode, which occurs at
the stagnation point along the azimuthal direction, is observed.

We examine the influence of the asymptotic speed alongside the scalar hair parameter on the dynamics of the
shock cone and the excitation of oscillation modes, uncovering intriguing results. As the absolute value of the
scalar hair parameter and the asymptotic speed both increase, the shock cone opening angle and the time required
for accreting matter to reach a steady state decrease. Concurrently, the stagnation point moves closer to the black
hole horizon in both scenarios. However, while the rest-mass density of the matter accreted inside the cone
decreases with an increase in the absolute value of $h/M$, it increases as the asymptotic speed rises. Thus, as
the hair parameter grows, the shock cone and QPOs gradually vanish, whereas the cone stability and the oscillation
frequencies are enhanced by an increase in the asymptotic speed.

In Fig. \ref{MA_a09_Diff_v}, we present the variation in the mass accretion rate at different asymptotic speeds
for $a/M = 0.9$ and $h/M = -0.25$, along with the Kerr solution at $V_{\infty}/c = 0.2$. At the same asymptotic
speed, the Horndeski solution for $h/M = -0.25$ significantly differs from the Kerr solution. This divergence
is evident both in the time it takes for the system to reach the steady state and in the pattern of matter accretion
towards the black hole. These differences influence the dynamics of the shock cone and the QPOs. Additionally,
variations in the asymptotic speed impact the stability and oscillations of the cone. Specifically,
at $V_{\infty}/c = 0.2$, the shock cone achieves stability without creating instabilities after reaching the
steady state, resulting in the absence of QPO formation in this model.

\subsection{The Comparison of $V_{\infty}/c =0.4$ from $a/M = 0.9$ with $h/M=-0.25$
  and $a/M = 0.6$ with $h/M=-0.8$}
\label{Compare_a09h025_a06h08}

To better understand the impact of asymptotic speed, we examine the structure of the shock cone and the
resulting QPOs in two black hole models: one with rapid rotation ($a/M = 0.9$) and the other with
moderate rotation ($a/M = 0.6$), both at the same asymptotic speeds ($V_{\infty}/c = 0.4$). We utilize
extreme hair parameters in both cases to deepen our understanding of the effects of asymptotic velocity
and the hair parameter on the formation of the shock cone in the vicinity of the Horndeski black hole.
The following results reveal that the asymptotic speed significantly influences the accretion mechanism,
the structure of the shock cone, and the QPOs. Indeed, to fully comprehend the impact of asymptotic speed
on these phenomena, further studies across a wide range of parameters are necessary. Through such research,
we anticipate that some observational results could be elucidated.

\begin{figure*}
  \vspace{1cm}
  \center
    \psfig{file=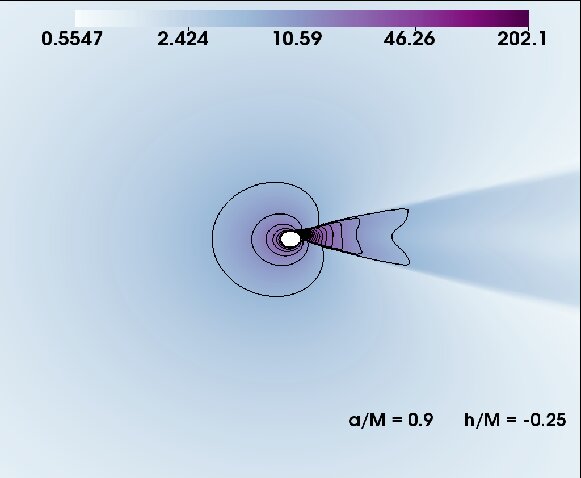,width=12.5cm}  
     \psfig{file=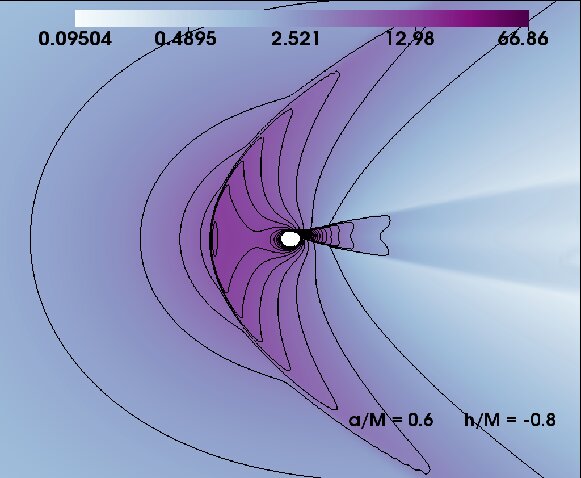,width=12.5cm}
     \caption{Same as in Fig.\ref{a06_color}  but 
       for $a/M = 0.9$ with $h/M=-0.25$  and $a/M = 0.6$ with $h/M=-0.8$ with the same
       asymptotic velocity $V_{\infty}/c =0.4$. The effect of the
       $V_{\infty}/c$ along with black hole scalar hair parameter is shown. }
\vspace{1cm}
\label{a09a06_color_Diff_v}
\end{figure*}

According to the results obtained from the numerical models with varying rotation and hair parameters,
it has been observed that the hair parameter significantly influences the dynamic structure of the formed
shock cone and the amount of matter trapped inside the cone. It can even cause the complete disappearance
of the shock cone and the trapped modes of QPOs. However, the black hole rotation parameter only alters
the value of the excited frequencies, without making a noticeable impact on the other conditions mentioned
above. Despite using different rotation parameters, as shown in Fig.\ref{a09a06_color_Diff_v}, it is numerically
observed that the rotation parameter does not affect the general physical structure of the cone. However,
as seen in Fig.\ref{a09a06_color_Diff_v} and other models, the scalar hair parameter, which significantly affects
the speed of matter falling towards the black hole asymptotically and spacetime, changes the formation of the shock cone.

In Fig.\ref{a09a06_color_Diff_v}, we modeled the dynamic structure of the shock cones in the presence of
strong scalar fields for two separate rotation parameters at the same asymptotic speed. According to
Fig.\ref{a09a06_color_Diff_v}, an increase in the intensity of the scalar field leads to changes in
the stagnation point within the cone and also results in a decrease in the amount of matter trapped
inside the cone. This, in turn, has led to the formation of a bow shock in the later stages (see the
bottom part of Fig.\ref{a09a06_color_Diff_v}). The bow shock is formed entirely by the matter on the
downstream side being pushed towards the upstream side due to the strong scalar field. The same phenomenon
is observed in the bottom left graph of Fig.\ref{a04_color}. However, in that case, since the asymptotic
speed is lower than the one used in Fig.\ref{a09a06_color_Diff_v}, the shock cone has completely disappeared,
and the matter has started to be pushed back in the direction from which it came. In other words, we can
only identify the formation of a bow shock. But since it is a weak bow shock, no QPO frequencies have
been observed in the numerical calculations.


\section{Possible QPO Models and Observed Frequencies from Numerical Simulations}
\label{Possible_QPO}
\begin{figure*}
  \vspace{1cm}
  \center
     \psfig{file=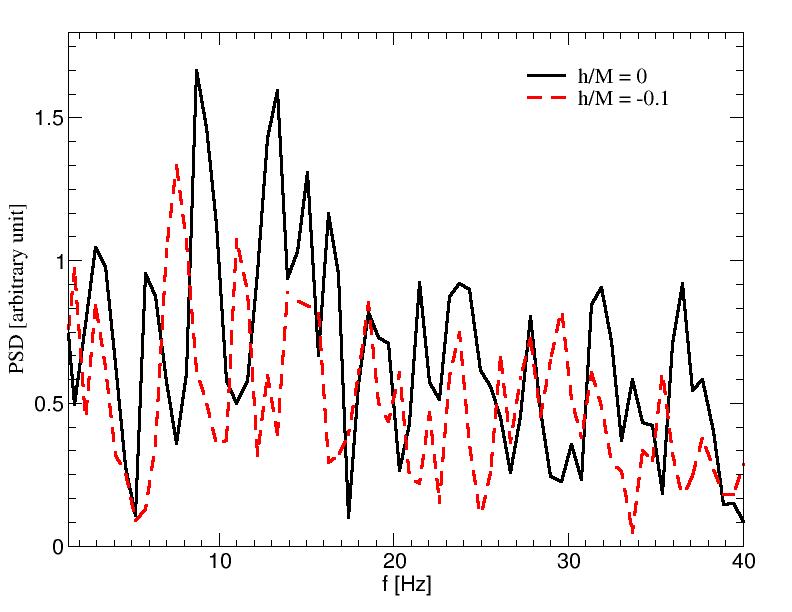,width=7.5cm}\hspace*{0.15cm}
     \psfig{file=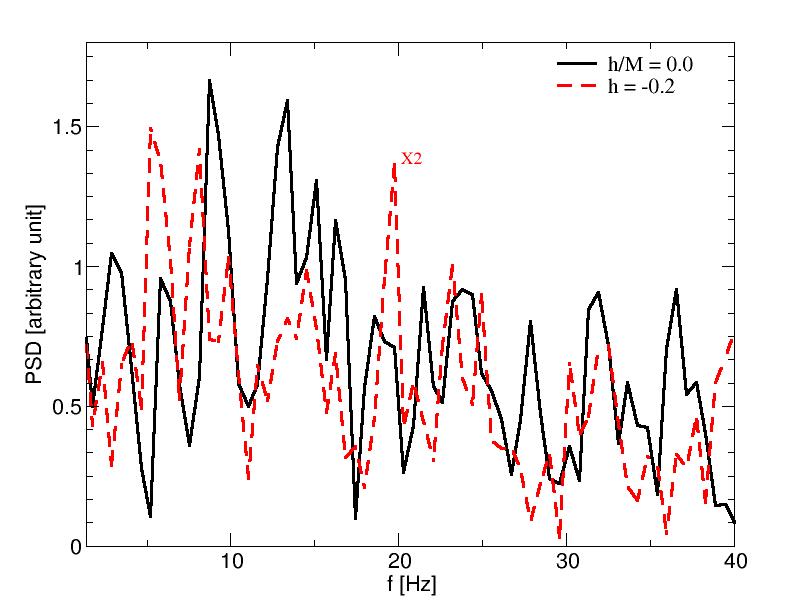,width=7.5cm}\\
    \vspace*{0.5cm} 
    \psfig{file=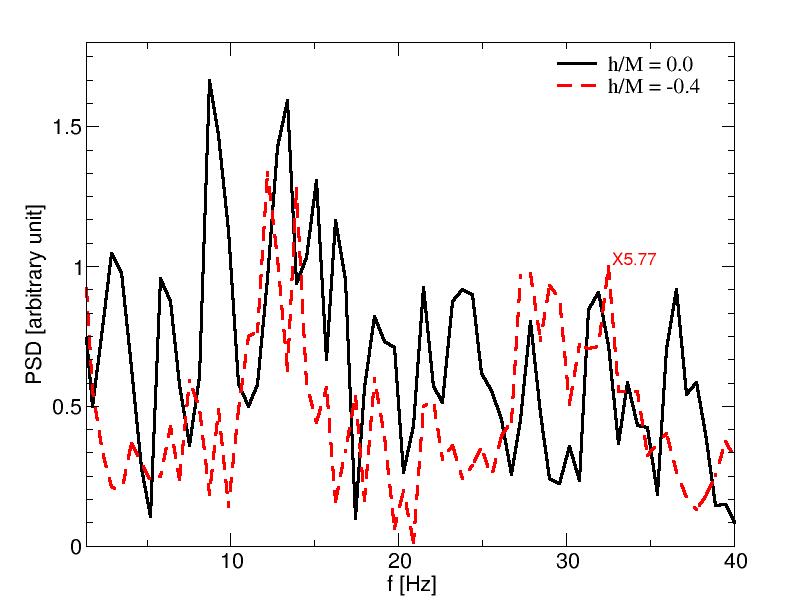,width=7.5cm}\hspace*{0.15cm}
    \psfig{file=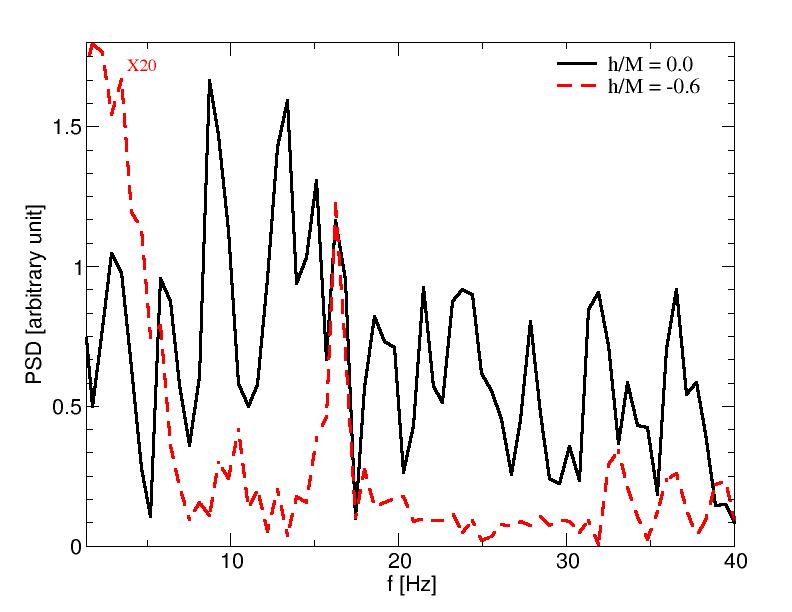,width=7.5cm}\\
    \vspace*{0.5cm} 
    \psfig{file=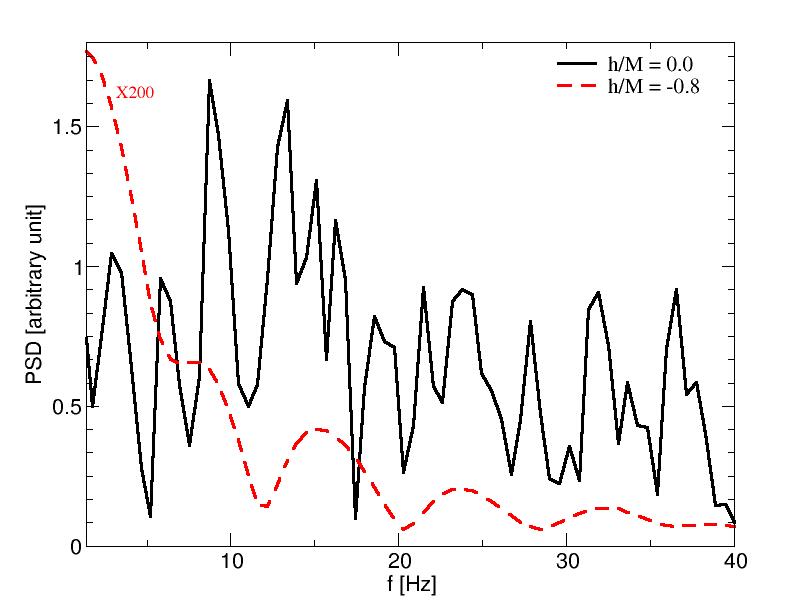,width=7.5cm}
    \caption{The Power Spectral Density (PSD) analysis has been calculated for different values of the hair
      parameter $h/M$ for $a/M=0.6$. PSD  analysis has been calculated from the  mass accretion rate data
      computed  near the black hole horizon. Since the value of $h/M=0$ produces the same solution as the
      Kerr black hole, by comparing the QPO frequencies at different values of $h/M$ with $h/M=0$,
      the effect of the scalar hair parameter on the formation of QPOs has been revealed.
      The mass of the black hole has been chosen as $M = 10M_\odot$.
      To directly compare the QPO oscillations in the Kerr case with oscillation frequencies in different hair
      parameter scenarios, the frequency amplitudes for different hair parameters have been multiplied by
      certain ratios. These are shown next to the red dashed line in each case.
}
\vspace{1cm}
\label{a06_PSD}
\end{figure*}

\begin{figure*}
  \vspace{1cm}
  \center
     \psfig{file=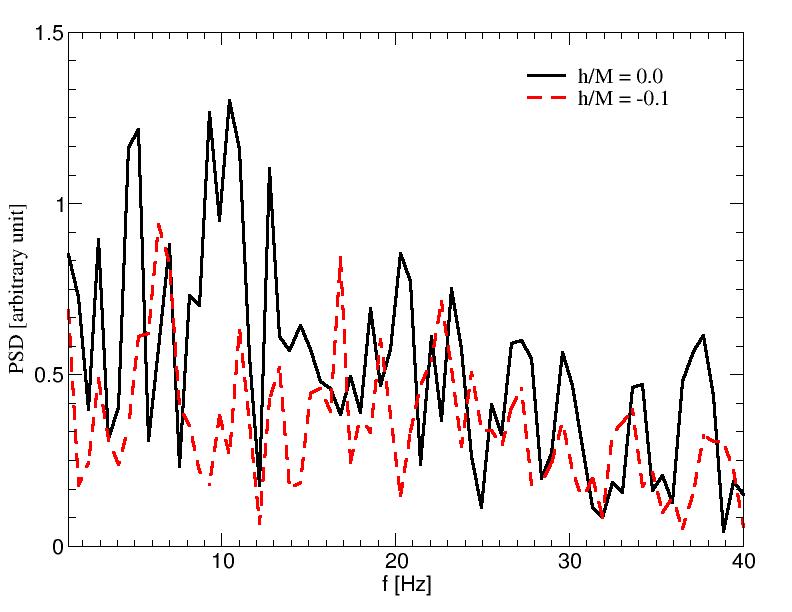,width=7.5cm}\hspace*{0.15cm}
     \psfig{file=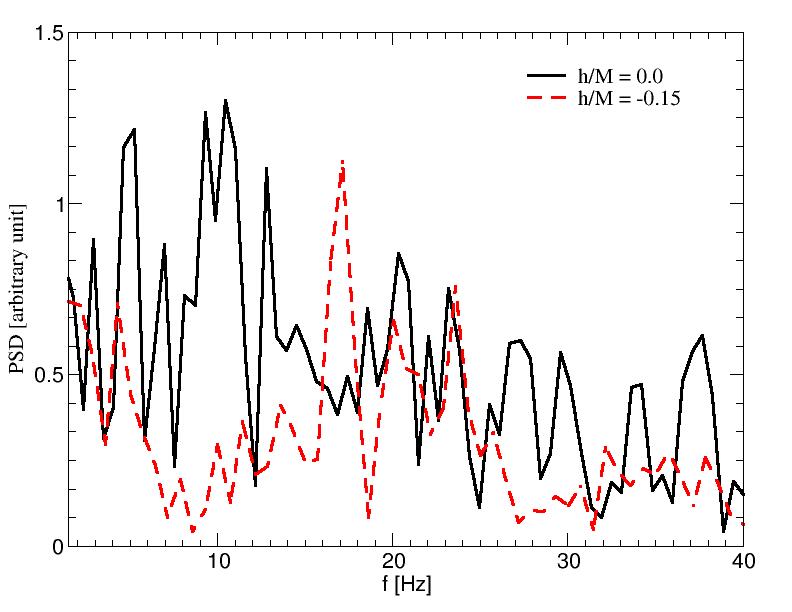,width=7.5cm}\\
    \vspace*{0.5cm} 
    \psfig{file=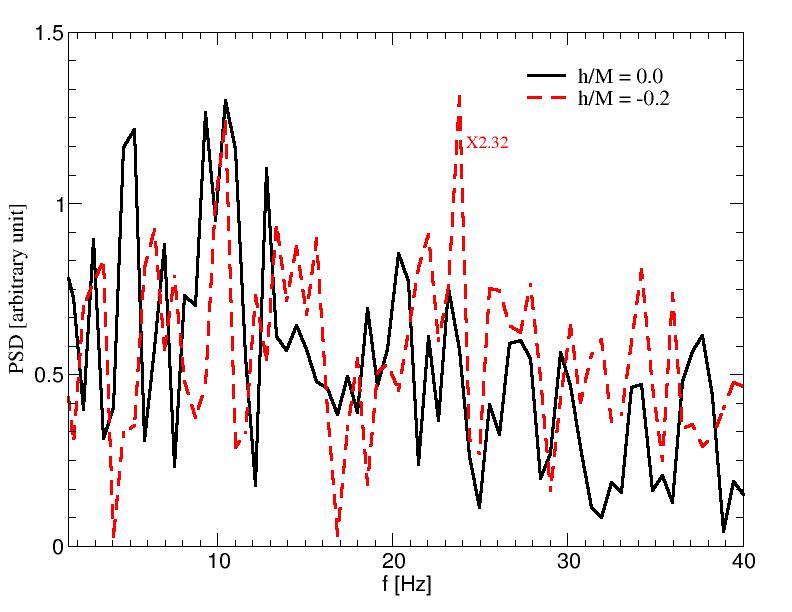,width=7.5cm}\hspace*{0.15cm}
    \psfig{file=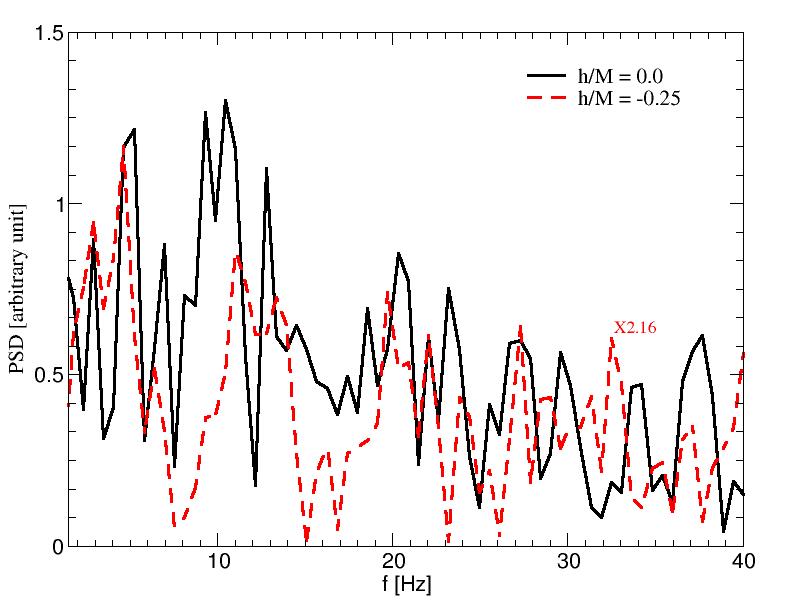,width=7.5cm}\\
    \caption{Same as in Fig.\ref{a06_PSD} but it is for $a/M=0.9$ models.
}
\vspace{1cm}
\label{a09_PSD}
\end{figure*}

Understanding the nature of QPOs through alternative theories of gravity
\cite{Maselli2015ApJ, Maselli2017ApJ, Rayimbaev2021, Rayimbaev2022PDU}
and an extra spatial dimension
\cite{Banerjee2017PhRvD, Banerjee2021JCAP}
could provide valuable insights for explaining some observational data in AGN and microquasar systems that
cannot be accounted for by Kerr gravity alone. In this context, we explore the oscillation
frequencies trapped by shock cones formed as a result of the BHL accretion around rotating
Horndeski black holes. Unlike Kerr black holes,
Horndeski black holes possess a scalar hair parameter, prompting us to investigate its effect on QPOs.
Within the shock cones, one of the intrinsic modes is the $p$ mode. The frequencies emerging from the
$p$ mode, formed within the torus and the shock cone around the Kerr black hole, depend on the black
hole rotation parameter and the accretion mechanism. However, the frequency of oscillations excited
by shock cones around the rotating Horndeski black hole not only depends on these parameters but
also undergoes significant changes due to the black hole hair parameter $h/M$. This variation
is numerically observed in the Power Spectral Density (PSD) analyses for $a/M=0.6$ and $a/M=0.9$,
as illustrated in Figs.\ref{a06_PSD} and \ref{a09_PSD}. This behavior is also reported
in Ref.\cite{CruzOsorio2023JCAP}.
The scalar hair parameter, which defines the scalar field as discussed in Ref.\cite{CruzOsorio2023JCAP},
not only modifies existing frequencies but also leads to the creation of new frequencies.
A schematic representation of these frequencies is provided in Figure 8 of Ref.\cite{CruzOsorio2023JCAP}.
As a consequence of the scalar hair parameter effect, various frequencies have been observed,
depending on the parameter. These include the oscillation frequency $f_{EH}$ between the stagnation point
and the black hole horizon at $r=2.3M$, the oscillation frequency $f_{\phi_{max}}$ between the stagnation
point and the point where the scalar field is maximum, the oscillation frequency $f_{bow}$ of the $p$
mode trapped along the azimuthal length, and the low-frequency $p$ mode oscillation $f_{sh}$,
as detailed in Ref.\cite{CruzOsorio2023JCAP}.

Based on the initial conditions given in Table \ref{Inital_Con}, as observed in Figs.\ref{a06_color}
and \ref{a09_color} through numerical modeling, no bow shock has formed. This reveals the presence of
three different fundamental modes, which are $f_{sh}$, $f_{EH}$, and $f_{\phi_{max}}$, in our numerical
simulations.
Fig.\ref{a06_PSD} demonstrates the variation of trapped QPO frequencies within the shock cone
formed on the downstream side of the rotating Horndeski black hole ($a/M=0.6$), depending on the
Horndeski hair parameter ($h/M$). Each plot compares Horndeski gravity under different $h/M$
(represented with a red dashed line) with Kerr gravity (represented with a black straight line),
revealing how QPO behavior changes with $h/M$. It is observed that the frequencies of the resulting
QPOs change; the fundamental modes gradually decrease and even fail to form for $h/M=-0.8$. This occurs
due to the decrease in the cone opening angle and the stagnation point, as calculated in numerical
simulations and provided in Table \ref{Inital_Con}. The stagnation point gets closer to the black hole
horizon as $h/M$ increases in the negative direction. The narrowing of the cone opening angle leads
to an increase in the $f_{sh}$ frequency. On the other hand, the stagnation point moving closer to the
black hole horizon causes the frequencies of the $f_{EH}$ and $f_{\phi_{max}}$ modes to increase and even disappear.

We compare the QPOs for every value of $h/M$ with the Kerr case. It is observed that while all QPO frequencies
occur in Kerr gravity, with different $h/M$ values, the amplitudes of the same frequencies decrease,
they appear at higher frequencies, or completely disappear for some models. For instance, at $h/M=-0.1$
and $h/M=-0.2$, these three fundamental QPO modes are formed. However, at $h/M=-0.4$, only two of them are
observed, and at $h/M=-0.6$, only one is detected. At $h/M=-0.8$, these three modes disappear. According to
our understanding from numerical observations, the two modes present at $h/M=-0.4$ are $f_{sh}$ and $f_{EH}$,
while the mode at $h/M=-0.6$ is $f_{sh}$. The disappearance of the other modes is attributed to the stagnation
point  proximity to the black hole. Apart from these modes, the peaks observed across all models result
from QPO frequencies generated by nonlinear coupling. $f_{sh}$, $f_{EH}$, and $f_{\phi_{max}}$ interact within
themselves, stacking on top of each other to create new QPO oscillation frequencies. These are also used
to explain the twin-peak QPOs \cite{Rayimbaev2023EPJC} or ratios such as $1:2:3...$ in their interpretation.

Fig. \ref{a09_PSD} demonstrates the behavior of QPO modes trapped within the shock cone
occurring in the downstream region around the Horndeski black hole with a rotation parameter
of $a/M=0.9$, as a function of $h/M$. As can be seen in Fig. \ref{a09_color} for $a/M=0.9$,
the structure of the shock cone formed around the black hole is similar across all $h/M$ values.
The only difference is in the amount of matter trapped within the cone, which decreases as $h/M$
becomes more negative. The reason for this is summarized in Table \ref{Inital_Con}, regarding the
behavior of the stagnation point and the opening angle of the cone. As $h/M$ increases in the negative
direction, the stagnation point moves closer to the black hole horizon. Simultaneously, the cone
opening angle decreases. However, since these changes are very small compared to the models with
$a/M=0.4$ and $a/M=0.6$, the overall structure of the shock cone is preserved. Therefore, it is
evident that the shock cone exhibits different behavior from the cases with $a/M=0.4$ and $a/M=0.6$.

Due to the preservation of the shock cone general dynamic structure and the stagnation point being
sufficiently far from the horizon, $f_{sh}$, $f_{EH}$, and $f_{\phi_{max}}$ have been observed to form within
the shock cone for each $h/M$. This situation is depicted in Fig.\ref{a09_PSD} alongside the Kerr
solution for different $h/M$ values. As shown in the figure, as $h/M$ changes, the resulting QPO
frequencies occur at higher values, while also displaying differences from the Kerr black hole.
As seen in Fig.\ref{a09_PSD}, the generated genuine modes lead to the formation of new frequencies
through nonlinear couplings. Both these genuine modes and their nonlinear couplings create resonance
conditions such as $3:2$, $5:3$, $2:1$, etc. This ratio also represents the lines where the nonlinear
couplings occur \cite{Remillard2002APS, Belloni2005A&A}.

As observed across all models of rotating black holes, the scalar hair parameter in Horndeski gravity
significantly influences the mass accretion rate, the formation of the shock cone, and the frequencies
of QPOs around the black hole. Therefore, the results from our numerical simulations could explain
observations that cannot be accounted for by Kerr gravity alone. This underscores the potential
importance of alternative theories of gravity, such as Horndeski gravity, in providing explanations
for astrophysical phenomena that deviate from the predictions of General Relativity (GR), as modeled
by the Kerr metric. The introduction of the scalar hair parameter unveils new dynamics that could help
bridge the gap between theory and observations, offering a more comprehensive understanding of the
universe most extreme objects.

\begin{figure*}
  \vspace{1cm}
  \center
  \psfig{file=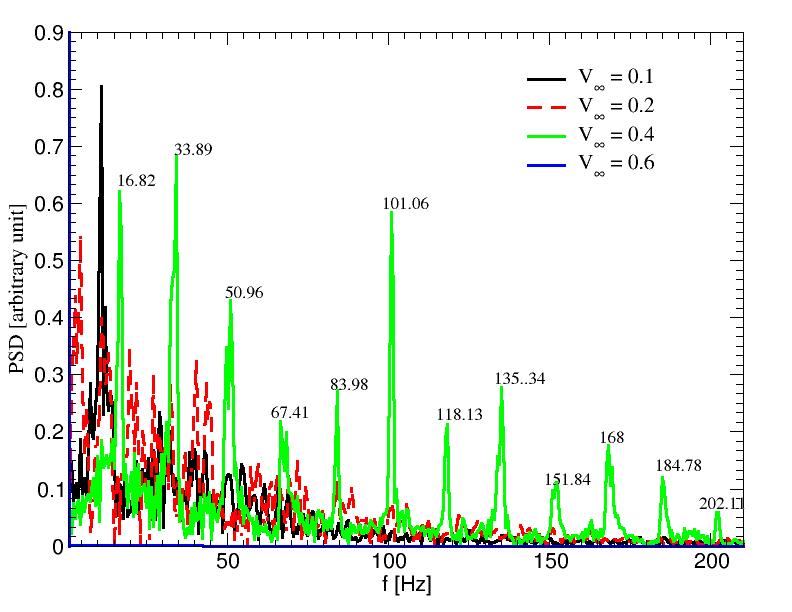,width=12.5cm}\\
  \vspace*{0.5cm} 
     \psfig{file=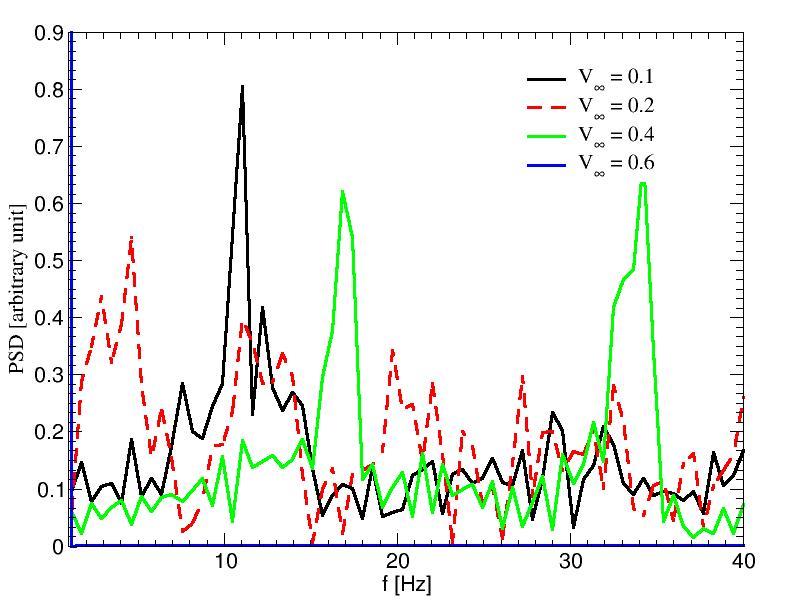,width=12.5cm}
     \caption{Same as in Fig.\ref{a06_PSD} but it is for $a/M=0.9$ with $h/M=-0.25$
       for various values of asymptotic velocity $V_{\infty}/c$.
       The top graph shows the PSD  results for each scenario across a wide frequency range,
       while the lower one displays changes up to only $40Hz$. In the case of the bottom PSD,
       comparisons at different asymptotic speeds can be made more clearly.
}
\vspace{1cm}
\label{a09_PSD_Diff_V}
\end{figure*}

In addition to the studies mentioned above, in cases where the hair parameter ($h/M=-0.25$) of the
rapidly rotating black hole ($a=0.9$) is close to the critical value, the behavior of the shock cone at
different asymptotic speeds reveals that the opening angle of the cone decreases, but the rest-mass density
within the cone increases. As shown in Fig.\ref{a09_PSD_Diff_V}, this situation significantly impacts the
frequencies derived from the numerical simulations.
In Fig.\ref{a09_PSD_Diff_V}, we examine the frequencies formed at different asymptotic speeds for
$a=0.9$ and $h/M=-0.25$. At $V_{\infty}/c=0.1$ and $V_{\infty}/c=0.2$, the frequencies $f_{sh}$, $f_{EH}$,
and $f_{\phi_{max}}$, along with their nonlinear couplings, are visible in the graph below Fig.\ref{a09_PSD_Diff_V}.
Interestingly, when the opening angle decreases and the stagnation point moves closer to the black hole for
$V_{\infty}/c=0.4$, the precise harmonic ratios of
$16.82:33.89:50.96:67.41:83.98:101.06:118.13:135.34:151.84:168:184.78:202.11 \equiv 1:2:3:4:5:6:7:8:9:10:11$ are
detected. This might suggest a highly structured and possibly resonant process
occurring in the accretion mechanism close to the black hole horizon.
From our understanding based on numerical simulation, it is concluded that as the stagnation point
approaches very close to the black hole horizon, all radial modes disappear. Only the $f_{sh}$ mode,
which results from the pressure mode, is formed, occurring at $16.8 Hz$ for the black hole with $M = 10M_\odot$.
Then, a series of frequencies produces the aforementioned ratio as a result of the nonlinear coupling of
this mode. This situation is clearly shown in the top graph of Fig.\ref{a09_PSD_Diff_V}.
Moreover, as the stagnation point moves closer to the horizon of the black hole for $V_{\infty}/c=0.6$, no
QPO frequency has been observed for $V_{\infty}/c=0.6$, as seen in both Figs.\ref{MA_a09_Diff_v} and
\ref{a09_PSD_Diff_V}. This indicates that, regardless of the shock cone solid structure or its
high rest-mass density, if the stagnation point approaches the horizon closer than a certain critical
value, it has been observed that no mode is excited within the cone.

\section{Possible Physical Mechanism and QPOs in M87*}
\label{M87}

$M87^*$ is a supermassive black hole located at the center of the NGC 4486 galaxy. It is a black hole
whose shadow was directly imaged by the Event Horizon Telescope (EHT)
\citep{Akiyama1, Akiyama2, Akiyama3, Akiyama4}, with a mass of approximately $6.5 \times 10^9 M_\odot$.
Following the first direct observation, numerous scientists have embarked on studies to further
understand the mass, spin parameters, and QPO frequencies of M87* \citep{Bandyopadhyay2022, Cui2023Nature}.
Additionally, various attempts have been made to elucidate the physical properties of these black holes
and their QPO frequencies using alternative theories of gravity.
One such alternative theory is Horndeski gravity. As discussed in detail in this paper, the spacetime
around a Horndeski black hole is influenced by both the spin and the scalar hair parameters. It has been
theoretically demonstrated that the spacetime experiences significant alterations compared to the Kerr
geometry, depending on the hair parameter. Consequently, Ref.\cite{Afrin2022ApJ} analytically applied
Horndeski gravity theory to the shadow of M87* as observed by the EHT, conducting a direct
parametric comparison.
In their theoretical comparison, they demonstrated that Horndeski gravity could account for the
characteristics of M87*. Assuming M87* is not a Kerr black hole but a Horndeski black hole, they
identified the ranges for the spin and hair parameters of the black
hole as $0.0077 \leq a/M \leq 0.9353$, $-0.756 \leq h/M < 0$ at $\theta_0=90$ (inclination angle),
and $0.0048 \leq a/M \leq 0.9090$, $-0.79 \leq h/M < 0$ at $\theta_0=17$.

In this paper, we examine the formation of shock cones during BHL accretion around rotating Horndeski
black holes, the dynamical behavior of these cones, and the excitation of QPO frequencies within the shock.
As detailed in Section \ref{Result1}, we consider cases such as $a/M=0.9$ with $h_{c}/M= -0.277$. When
$\lvert h/M \rvert < \lvert h_{max}/M \rvert$, we model and observe shock cone formation across various
models. Given the role of shock cones in exciting QPO modes, we have observed the presence of QPO frequencies
for each model at $a/M=0.9$, including genuine modes and their nonlinear couplings.
For $a/M=0.6$, with $h_{max}/M$ at $-0.983$, we explore hair parameters for modeling up to
$\lvert h_{max}/M \rvert > \lvert h/M \rvert =0.8$. We note the formation of shock cones and QPO
frequencies again, with significant changes in the shock cone physical dynamics at $h/M=-0.6$ and $h/M=-0.8$.
The change at $h/M=-0.6$ is less pronounced, allowing the excitation of some modes but not all, resulting in
observed QPO frequencies. At $h/M=-0.8$, the shock cone structure significantly deforms, nearly reaching
the phase of disappearance, leading to no observed QPO frequencies.
Lastly, at $a/M=0.4$, with $h_{max}/M=-1.367$, the range of $h/M$ is notably broad. Numerical simulations
have shown that at $\lvert h/M \rvert \geq 0.8$, the shock cone dynamical structure either changes completely
or disappears, preventing the excitation of QPO modes. However, for $\lvert h/M \rvert < 0.8$, both the
cones and QPO frequencies form.

As a result, the ranges of $h/M$ and $a/M$ theoretically defined for $M87^*$, as outlined in the
previous paragraph for different inclination angles, directly align with the values of $h/M$
necessary for the excitation of shock cones and QPO modes in our numerical models. Thus, we can
propose the BHL scenario as the formation mechanism for the shock cone around the $M87^*$ black hole.
These shock cones are candidates for a physical mechanism that could excite QPO frequencies, although
these frequencies have not yet been observed in this source.

While no specific QPO frequencies have been identified from $M87^*$ observations,
both theoretical and numerical models have suggested potential QPO frequencies by
describing the spacetime fabric around the M87 black hole with a scalar hair parameter from studies.

The QPO frequencies have not yet been detected in observations of $M87^*$. Theoretical studies
\cite{Afrin2022ApJ} have defined the hair parameter for gravity around $M87^*$. This paper demonstrates,
through numerical analysis, that the potential scalar hair parameters for $M87^*$ are consistent with theory.
The agreement between theory and numerical simulations suggests that the shock cone physical mechanism
proposed here for the $M87^*$ source could be responsible for the potential QPOs observable from this
source. Based on these models and corresponding hair parameters, we discuss the possible QPO
frequencies that could be observed from the $M87^*$ source.

The PSD analyses in Figs. \ref{a06_PSD} and \ref{a09_PSD}, recalculated for the mass of the $M87^*$
black hole, $M = 6.5 \times 10^9 M_\odot$, reveal that the frequencies resulting from both genuine
modes and their nonlinear couplings could occur in the range of $4.6 \times 10^{-9} - 1.53 \times 10^{-6}$ Hz.
As discussed in Chapter \ref{Possible_QPO}, the characteristics of the observed frequencies can be
said to fall within a certain range, entirely dependent on the hair and black hole spin parameters.
It has been observed that as the hair parameter increases in the negative direction, the frequency also increases.


\section{Understanding the Horndeski parameter-Mass-QPO Relation in
  GRS $1915+105$}
\label{Horndeski_Mass_QPO}
GRS 1915+105 is one of the well-known $X$-ray binaries in our galaxy, exhibiting remarkable $X$-ray
variability that has drawn significant attention. Analytical and observational studies have shown that
the mass of the black hole at the center of this microquasar is approximately $M \sim 12.5M_\odot$.
Observations suggest that the QPO frequencies vary within the range of $1-10$ Hz
\cite{Rawat2019ApJ, Misra2020ApJ, Tian2023Natur} and include $34$ Hz and $67$ Hz in the X-ray band
\cite{Belloni2013MNRAS}. It is believed that these oscillations originate from the inner radius of the
computational domain \cite{Misra2020ApJ, Chauhan2024MNRAS}. In this section, we attempt to establish
the relationship between the mass of the black hole and QPOs with the scalar hair parameter $h/M$ by
comparing our numerical results, which are derived from the mass accretion rate calculated at the
inner radius of the computational domain.

When the PSD analysis shown in Figs. \ref{a06_PSD}, \ref{a09_PSD}, and \ref{a09_PSD_Diff_V} for the
black hole GRS 1915+105 is recalculated, it is observed that the frequencies vary between $2.3$ Hz and
$f_{nm}$. Here, $f_{nm}$ varies depending on the black hole rotation parameter, the asymptotic velocity
of the gas in BHL accretion, and the scalar hair parameter $h/M$. $f_{nm}$ could reach up to $\sim 500$ Hz.
When considering $h/M$, the observed frequencies of the source occur only when $\lvert h/M \rvert < 0.25$
for $a/M=0.6$ or $a/M=0.9$. If we consider the situation based solely on the asymptotic velocity $V_{\infty}/c$,
agreement with observations occurs when $V_{\infty}/c < 0.4$. On the other hand, for the observational QPOs to
be consistent with a scenario where $\lvert h/M \rvert \geq 0.25$, or for the asymptotic velocity to be at
$V_{\infty}/c = 0.4$, the mass of the GRS 1915+105 black hole must be in the range of $12M_\odot \leq M < 400M_\odot$.


\section{Discussion and Conclusion}
\label{Conclusion}
We have studied the impact of the scalar field, as defined by Horndeski gravity, on the shock cone
formation around rotating black holes. This scalar field, also known as the hair parameter of the
Horndeski black hole, modifies the gravitational potential of the spacetime, thereby affecting the
mass accretion rate around the black hole. We demonstrate how the hair parameter influences spacetime
and, in turn, alters the physical structure of the shock cone resulting from BHL accretion. With
changes in the cone structure, we identify conditions under which pressure-based and radial-based
QPO frequencies are either excited through nonlinear coupling or completely disappear. These
findings underscore the effects of parameters such as the black hole rotation, the scalar
field hair parameter, and the asymptotic velocity of the matter injected from the outer boundary.

The scalar hair parameter in Horndeski gravity has different critical values depending on the black hole
rotation parameter. In the model of a rapidly spinning black hole with $a/M=0.9$, the numerical results
of the shock cones in the vicinity of the Horndeski black hole are similar to the results of Kerr ones
because the absolute maximum value of the scalar hair parameter is not very large. Nonetheless, the
dynamics of QPO oscillations have been influenced due to the alteration in how spacetime
interacts with the scalar potential under varying scalar hair values. As $h/M$ approaches the
critical value, the frequency of the observed fundamental mode has increased. The same situation
is also true for slowly rotating black holes. Since the absolute critical hair parameter can take
on possibly large values in the slowly rotating black hole cases, the influence of the strong scalar
field on the gravitational potential has been stronger. Consequently, as $h/M$ approaches the critical
value, a significant change in the physical structure of the shock cone has been observed. Also, the
opening angle of the shock cone has decreased, the stagnation point has approached closer to the black
hole horizon, and the cone has reached the steady state more quickly. At the same time, as more matter
within the cone begins to move away from the black hole, the shock cone has started to disappear entirely.
Even for $a/M=0.4$ with $h/M =-1.2$, all the matter in the region of the shock cone is expelled outward due
to the potential of the scalar field. Because of these physical changes, not only have the behaviors and
frequencies of the QPOs excited within the shock cone changed, but in the case of a strong scalar field, the
QPO frequencies have completely disappeared.

In addition to the scalar hair parameter in Horndeski gravity, modeling some special cases for the
asymptotic velocity has revealed the effect of this velocity, along with the hair parameter, on the
physical structure of the resulting shock cone and on QPOs.
In the case of $V_{\infty}/c<0.4$, it is observed that all modes within the shock cone are excited,
including the fundamental oscillation modes and their nonlinear couplings. However, at $V_{\infty}/c=0.4$,
only the $f_{sh}$ frequency, created due to the pressure mode, is excited, and this frequency has been
seen to exhibit a perfect harmonic resonance. For $V_{\infty}/c>0.4$, no oscillation mode is found. This
confirms that the asymptotic velocity, together with the hair parameter, significantly affects both
the structure of the resulting shock cone and the excited QPOs.
In our future work, we plan to model different values of $V_{\infty}/c$ with different rotation
and hair parameters, thereby revealing the physical mechanisms that these three different physical
parameters can create. Thus, we will be able to uncover the physical mechanisms behind different
observational results from X-ray binaries and AGNs.

The effect of the asymptotic speed and the scalar hair parameter on the dynamics of the shock cone and
the excitation of oscillation modes is revealed. The investigation uncovers that an increase in both
the absolute value of the scalar hair parameter and the asymptotic speed leads to a reduction in the
cone opening angle and a shortened duration for the cone to achieve steady-state. However, the stagnation
point approaches the black hole horizon in both scenarios.
While the rest-mass density of the matter accreted inside the cone diminishes with an increase in
the absolute value of $h/M$, this density escalates with an increase in the asymptotic speed. Therefore,
an increase in the scalar hair parameter in the negative direction can cause the gradual disappearance
of the shock cone and QPOs, while the stability of the cone increases with the asymptotic speed,
also altering the oscillations.

The formation of shock cones resulting from BHL accretion, depending on the hair
parameter, and the excited QPO frequencies within these cones have also been revealed. Numerical
results have shown that shock cones occur at certain values of $h/M$, especially for models of
slowly rotating black holes. These $h/M$ values are in agreement with the analytical work conducted
by Ref. \cite{Afrin2022ApJ} on the observed $M87^*$ by the Event Horizon Telescope (EHT). Therefore,
the shock cones and the QPOs excited within these cones found in our numerical studies could be
suggested for the $M87^*$ source.
Additionally, based on observational results from the GRS 1915+105 black hole, the possible scalar
hair parameter for this black hole has been defined. It has been determined that $h/M > -0.25$ and
$V_{\infty}/c < 0.4$ are required. On the other hand, for the frequencies obtained from observations
to occur with other models used in numerical simulations, $h/M \leq -0.25$ and $V_{\infty}/c = 0.4$.
It has been concluded that the mass of the black hole must be $12 M_\odot \leq M \leq 400 M_\odot$.

In summary, the interaction of weak and strong scalar fields with spacetime has revealed changes in the
dynamic structure of shock cones formed around stationary rotating Horndeski black holes during BHL
accretion and the behavior of QPO frequencies excited within the cone. The numerical results we
have obtained here might be used to provide solutions to some observational data that are not
explained by Kerr gravity. Simultaneously, they can be employed to address some mysteries of the
universe, such as why QPOs are not observed from some sources. For instance, due to their characteristic
structures or certain observational difficulties, QPO behaviors have not been fully determined
in XTE J1550-564 \cite{Rink2022MNRAS} and GX 339-4 \cite{Zhang2023, Zhang2024MNRAS}.
Our numerical results could offer an explanation for such X-ray systems.

Superradiance is a phenomenon that occurs as a result of the interaction between a black hole and the surrounding matter. Although the GRH equations we use do not include radiation terms, we can still reveal superradiance scenarios \citep{Cardoso2004PhRvD, Brito2015SuperradianceEE}. To do this, after forming a shock cone around the black hole and reaching a steady state, the stability of the shock cone can be examined by perturbing one of the parameters such as density, velocity, or scalar field. The response of the shock cone to the perturbation is calculated over time. Changes in the angular momentum transfer of the cone and the parameters defining the cone are numerically observed in the strong gravitational region, allowing us to study the evolution of the gravitational and scalar fields. However, since the main purpose of the paper is to reveal the impact of the Horndeski scalar field on the shock cone and QPOs, these situations will be addressed in future planned studies.


\section*{Acknowledgments}
All simulations were performed using the Phoenix  High Performance Computing facility at the American University of the Middle East (AUM), Kuwait. We sincerely thank the referee for their significant contributions to improving the paper and for shedding light on a potential future study.\\

\bibliographystyle{JHEP} 
\bibliography{paper.bib}

\end{document}